\documentclass[fleqn,10pt]{wlscirep}
\usepackage[utf8]{inputenc}
\usepackage[T1]{fontenc}
\usepackage{lineno}
\usepackage{pifont}
\usepackage{xcolor}
\usepackage{tabularx}
\usepackage{hyperref}       
\usepackage{url}            
\usepackage{booktabs}       
\usepackage{amsfonts}       
\usepackage{nicefrac}       
\usepackage{microtype}      
\usepackage{array}
\usepackage{comment}
\usepackage{rotating}
\usepackage{wrapfig}
\usepackage{float}
\usepackage{microtype}
\usepackage{graphicx}
\usepackage{booktabs} 
\usepackage{caption}
\usepackage{multirow}
\usepackage{ulem}
\usepackage{enumitem}
\usepackage{amsmath}
\usepackage[noend]{algorithmic}
\usepackage{algorithm}
\usepackage{algorithmic}
\usepackage{subcaption}
\usepackage{bm}
\usepackage{marvosym}
\usepackage{url}
\usepackage{hyperref}

\newcolumntype{Y}{>{\centering\arraybackslash}X}

\title{WorldMove, a global open data for human mobility}

\author[1,2,*]{Yuan Yuan}
\author[1,2,*]{Yuheng Zhang}
\author[1,2]{Jingtao Ding}
\author[1,2,\dag]{Yong Li}
\affil[1]{Department of Electronic Engineering, Tsinghua University, Beijing, P.R. China}
\affil[2]{Beijing National Research Center for Information Science and Technology (BNRist), Beijing, P.R. China}
\affil[*]{Equal contribution}
\affil[$\dag$]{Corresponding author(s): Yong Li (liyong07@tsinghua.edu.cn)}

\begin{abstract} 
High-quality human mobility data is crucial for applications such as urban planning, transportation management, and public health, yet its collection is often hindered by privacy concerns and data scarcity, particularly in less-developed regions.
To address this challenge, we introduce WorldMove, a large-scale synthetic mobility dataset covering over 1,600 cities across 179 countries and 6 continents.
Our method leverages publicly available multi-source data, including gridded population distribution, point-of-interest (POI) maps, and commuting origin-destination (OD) flows, to generate realistic city-scale mobility trajectories using a diffusion-based generative model. The generation process involves defining city boundaries, collecting multi-source input features, and simulating individual-level movements that reflect plausible daily mobility behavior. 
Comprehensive validation demonstrates that the generated data closely aligns with real-world observations, both in terms of fine-grained individual mobility behavior and city-scale population flows.
Alongside the pre-generated datasets, we release the trained model and a complete open-source pipeline, enabling researchers and practitioners to generate custom synthetic mobility data for any city worldwide. 
WorldMove not only fills critical data gaps, but also lays a global foundation for scalable, privacy-preserving, and inclusive mobility research, empowering data-scarce regions and enabling universal access to human mobility insights.
\end{abstract}
\begin{document}

\flushbottom
\maketitle

\thispagestyle{empty}


\section*{Background \& Summary}

High-quality human mobility data plays a crucial role in diverse domains such as urban planning~\cite{zheng2023spatial,zheng2025urban}, transportation management~\cite{yuan2021survey,yuan2024unist}, public health interventions~\cite{yuan2022activity,feng2020learning}, and sustainable development~\cite{zeng2024estimating}.
However, recent research has revealed that existing mobility datasets, which are often derived from isolated platforms or providers, can exhibit significant structural biases. For instance, Gallotti \textit{et al.}~\cite{gallotti2024distorted} demonstrate that different mobility data sources (e.g., GPS, CDR, census, Google, Facebook) show substantial divergences in fundamental patterns such as movement distance distribution and network connectivity, which in turn can lead to contradictory outcomes in downstream modeling tasks. These discrepancies are particularly problematic when such data are used as universal proxies for human mobility without adequate scrutiny or cross-validation.
In parallel, data access remains severely constrained due to privacy concerns, proprietary restrictions, and infrastructural limitations, especially in low-resource regions. Consequently, most large-scale studies are skewed toward well-surveyed cities or countries, exacerbating global knowledge asymmetries and limiting the universality of urban and behavioral research.

To address these challenges, there is a growing interest in using generative artificial intelligence (AI) to produce synthetic mobility data. Recent works~\cite{yu2023city,li2023learning,yuan2024generating,yuan2022activity} have successfully employed AI models to simulate trajectory data by learning from limited samples. 
For example, Feng \textit{et al.}~\cite{feng2020learning} and Yuan \textit{et al.}~\cite{yuan2023learning} propose GAN-based models to simulate human mobility from real-world data. Similarly, diffusion models~\cite{zhang2024noise,zhu2023difftraj} have been employed to generate synthetic mobility trajectories.
However, these efforts are typically constrained to limited cities and often require access to real-world trajectory data for each target region, thus failing to overcome the global coverage gap or alleviate dependence on city-specific ground truth data.

\begin{figure}[h]
\vfill
\centering
\includegraphics[width=\linewidth]{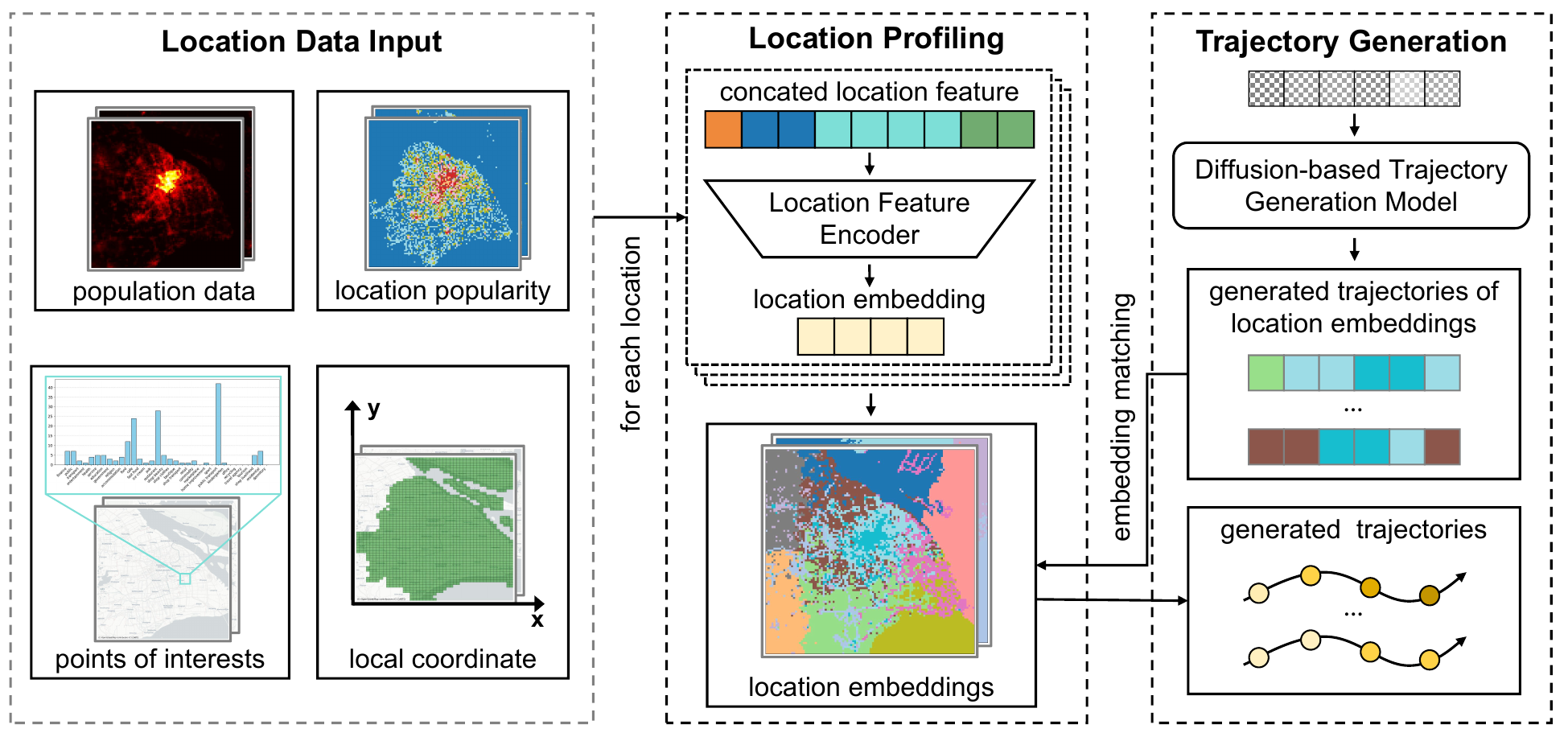}
\caption{Overall framework of the mobility generation system.}
\label{fig:method_flow}
\end{figure}

In this paper, we introduce \textbf{WorldMove}, a large-scale synthetic mobility dataset generated via a diffusion-based generative AI model trained on multi-source real-world data, including GPS, CDR, and credit card-based mobility records. 
\color{black} 
Our method explicitly acknowledges the structural biases across different data sources and leverages their complementary strengths by first abstracting all locations into a unified, data-agnostic embedding space. By training the generative model within this common semantic space, we produce a harmonized model of mobility that learns the fundamental patterns of movement independent of the original data source, thereby producing realistic synthetic mobility data across 1600+ cities worldwide.
The generation pipeline combines geographic features (population, POIs) and mobility semantics (OD flows) within a shared embedding space\color{black}, enabling model transferability and generation for cities with no available mobility data. As illustrated in Figure~\ref{fig:method_flow}, our generation process consists of three steps: (1) encoding multi-source location features into a unified embedding space; (2) training a diffusion model on real mobility data to generate trajectories in this space; and (3) mapping the generated embeddings to the target city's locations via minimum-distance matching to construct the final trajectories.

Our contribution is twofold: (1) we release a globally unified and privacy-preserving mobility dataset that can serve as an open resource for large-scale urban and social analysis; and (2) we provide an extensible generation pipeline and pretrained model that allow practitioners to generate synthetic trajectories for any city of interest. Through extensive evaluations, including distributional similarity, aggregate OD matching, mobility law recovery, geospatial realism, and privacy protection, we demonstrate that the data generated by WorldMove is highly realistic, effectively captures key human mobility patterns, and preserves user privacy.
Moreover, we showcase the utility of WorldMove through two representative applications: simulating urban transportation emissions for sustainable mobility planning and modeling green space exposure to support public health and equity research.
Table~\ref{tbl:cmopare_data} highlights the advantages of our dataset over existing real-world and synthetic mobility datasets.
By addressing critical challenges such as limited data availability and the lack of accessible mobility datasets, WorldMove lays a solid foundation for advancing fair, transparent, and reliable urban modeling and policy simulation on a global scale.

\section*{Methods}

\begin{figure}[p]
\centering
\includegraphics[width=\linewidth]{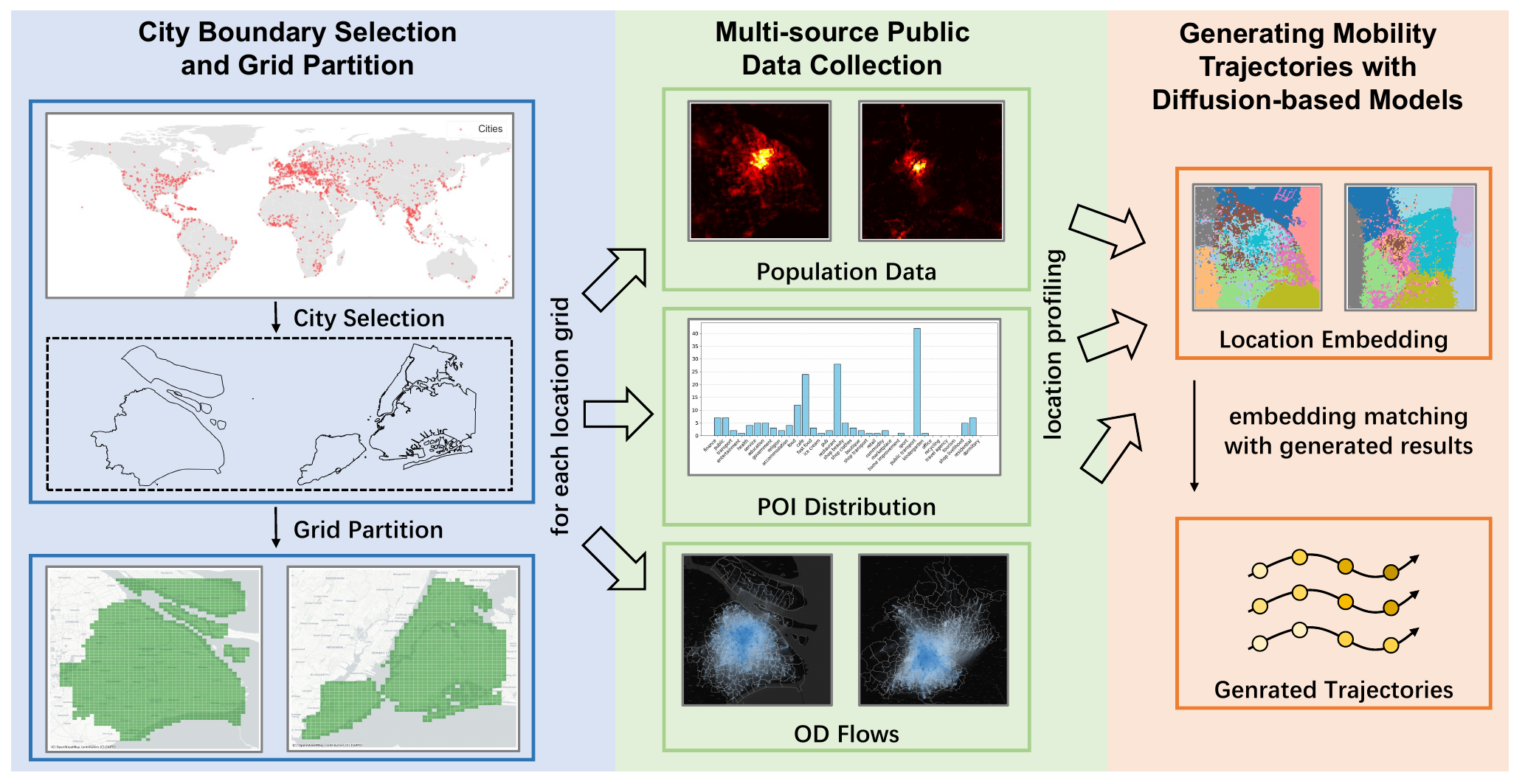}
\caption{Overview of the dataset construction pipeline. (1) Determining city boundaries and geographic units (grids). (2) Multi-source public
data collection (population, points of interest, od flow). (3) Generating mobility trajectories with diffusion-based models.}
\label{fig:data_flow}
\end{figure}

In this section, we present the methodology for constructing the WorldMove dataset. As illustrated in Figure~\ref{fig:data_flow}, the pipeline consists of three main stages. First, we define city boundaries and partition each city into locations. Second, we extract features from multi-source global public data to learn location representations. Finally, we employ diffusion-based generative models to synthesize realistic mobility trajectories.

\color{black}

\subsection*{Problem definition}

We aim to generate realistic synthetic mobility trajectories for cities by learning from publicly available spatial and functional data. The problem can be defined in terms of its inputs and outputs as follows.

\noindent \textbf{Inputs:}
\begin{itemize}[leftmargin=*]
    \item \textbf{City boundaries}: The geographical extent of the target city that is used to define its spatial scope.
    \item \textbf{Multi-source global public data}: It includes population density, points of interest (POIs), location popularity, local coordinates, and origin-destination (OD) flows, providing spatial context and functional attributes of the city.
\end{itemize}

\noindent \textbf{Outputs:}
\begin{itemize}[leftmargin=*]
    \item \textbf{Location representations}: Learned embeddings of each partitioned area within the city, capturing both spatial structure and functional characteristics.
    \item \textbf{Synthetic mobility data}: Realistic trajectories that reproduce plausible individual movements and aggregate population flows within the city.
\end{itemize}

The input data used for our model, including gridded population distribution, point-of-interest maps, and commuting flows, were derived from publicly available and aggregated sources. Our final generated dataset, WorldMove, is purely synthetic and contains no individually identifiable human subject data. Therefore, this study was not subject to oversight by an Institutional Review Board (IRB) and did not require IRB approval.

\color{black}
\subsection*{Defining city boundaries and locations}

\begin{figure}[p]
\centering
\includegraphics[width=\linewidth]{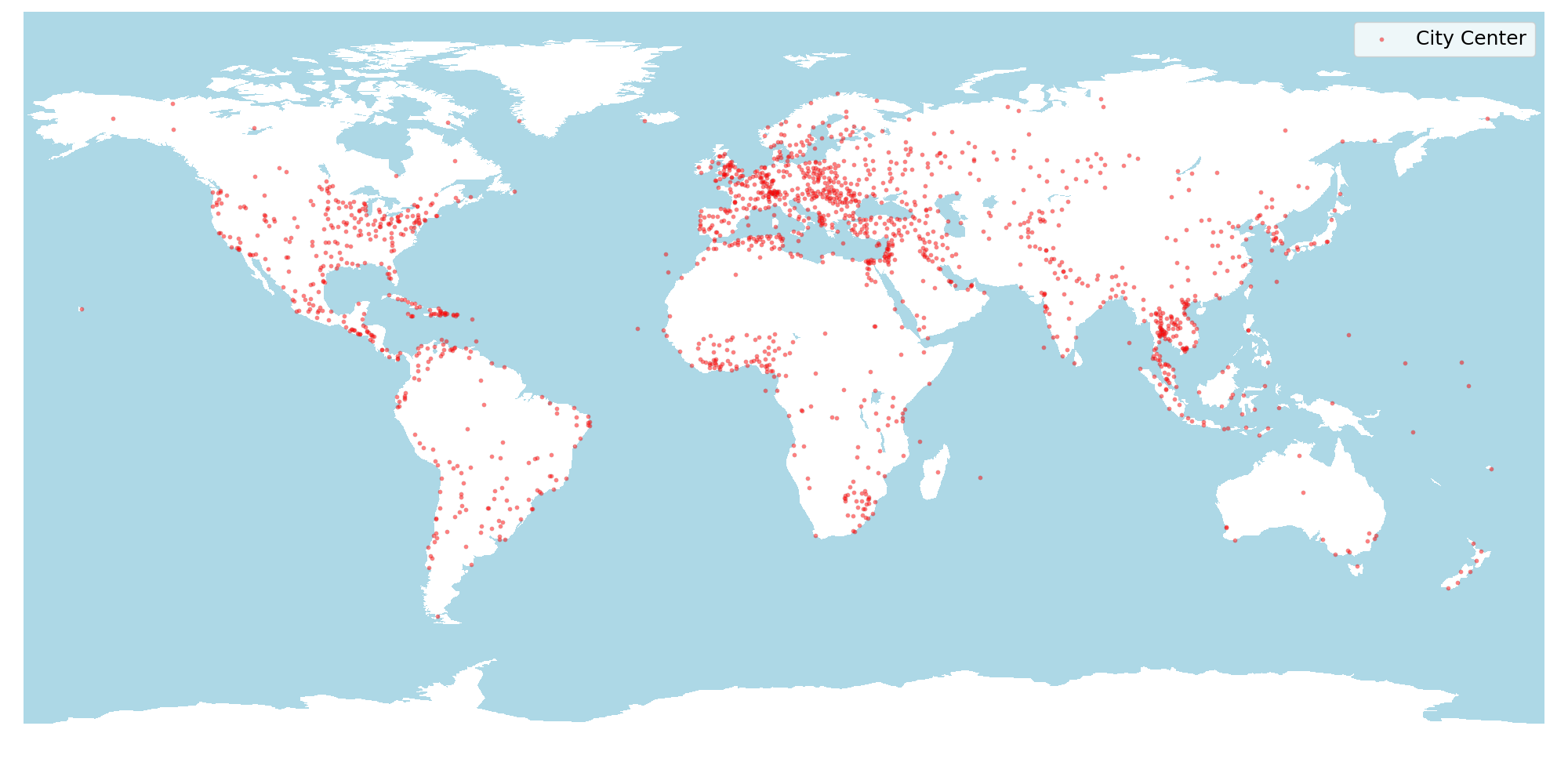}
\caption{An overview of the globally distributed cities included in the dataset.}
\label{fig:global}
\end{figure}

For each city, we begin by defining its geographic boundaries using map tools to ensure an accurate representation of its spatial extent. To achieve comprehensive global coverage, we integrate data from multiple sources to establish detailed and reliable boundaries for cities. Specifically, we utilize the WhosOnFirst (WOF) database (\url{https://whosonfirst.org/}), an open-source global gazetteer of places that provides boundary data in GeoJSON format. From this dataset, we select cities with explicitly defined boundaries represented as polygons, ensuring consistency and precision for each selected city. Figure~\ref{fig:global}  shows the globally distributed cities included in our WorldMove dataset.

\begin{figure}[p]
\centering
\includegraphics[width=\linewidth]{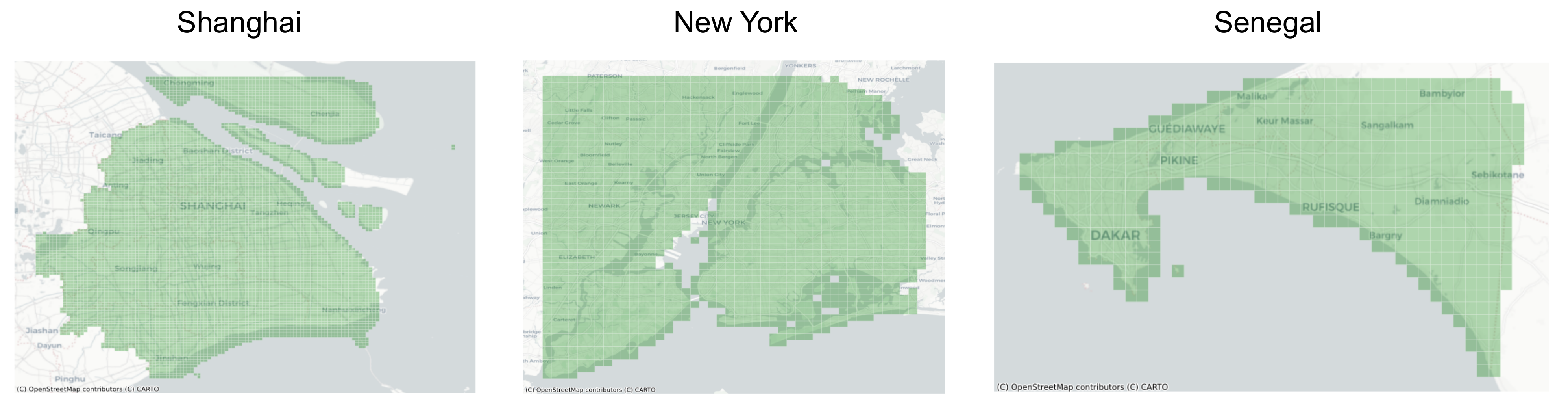}
\caption{Example of the region division for three cities.}
\label{fig:division}
\end{figure}

Once the city boundary is defined, the area is divided into uniform grids, with each grid cell measuring 1 km × 1 km, as shown in Figure~\ref{fig:division}, which illustrates examples of the regional division. 
\color{black}
This resolution was chosen because it strikes a balance between location precision and computational efficiency. It captures fine-grained mobility patterns across both high-traffic areas (e.g., commercial districts) and less frequently visited zones. This 1 km grid size is widely adopted in mobility studies~\cite{li2023learning,long2025universal,zhang2024noise} as it provides a detailed representation of urban environments.
Larger grid sizes, such as 5 km × 5 km, would fail to capture the spatial nuances of human movement.
\color{black}
These grids form the foundational units of "locations" in our mobility trajectory data. It is important to note that a "location" does not represent a specific point but rather a small, bounded area within the city.

\subsection*{Globally accessible multi-source data for location profiling}
\label{sec:2.2}

To create a comprehensive profile for each location, we integrate various globally accessible multi-source datasets that provide key attributes for every grid cell. The profiling incorporates the following components:

\begin{itemize}[leftmargin=*]
    \item \textbf{Population data:}  We use population data from the WorldPop dataset (\url{https://www.worldpop.org/}), which provides high-resolution (100m) population distribution estimates. This data enables an accurate representation of population density within each grid in the selected city area. 
    
    \item \textbf{Points of interests (POIs):} POIs effectively capture urban functional characteristics as they represent specific sites associated with various human activities. In our dataset, POI data is sourced from OpenStreetMap (OSM) (\url{https://www.openstreetmap.org/})  and categorized by type within the boundaries of each urban region. This process generates a 34-dimension vector that reflects the distribution of POIs by type, providing a detailed description of the functional attributes of the corresponding area.
    
    \item \textbf{Location popularity:} In addition to the static characteristics of each location, such as population density and POI distribution, we also extract mobility-related features, specifically location popularity. This is quantified as the visitation frequency rank, derived from a global high-resolution origin-destination commuting flow dataset (\url{https://fi.ee.tsinghua.edu.cn/worldod/}). By analyzing this dataset, we calculate the visitation flow for each location, providing a movement perspective on its significance within the mobility network.
    
    \item \textbf{Local coordinate system:} For each location, apart from the features related to trajectory semantics, we also introduce a local coordinate system to assist our model in learning the spatial relationships between the locations visited by the trajectory. As shown in Figure~\ref{fig:method_flow}, for each selected city's boundary, a two-dimensional coordinate system is established based on the minimum bounding rectangle that encloses the boundary. Each location within this boundary is assigned a two-dimensional coordinate, which is then normalized to the range of 0 to 1 using the rectangle’s length and width as the maximum values. The incorporation of local coordinates aids the model in understanding the relative distances between regions, thereby facilitating the generation of more realistic mobility data.
    
\end{itemize}

Combining the above four aspects, we construct a feature vector for each location, resulting in a vector of 38 dimensions: 1 dimension derived from population data, 34 from points of interest, 1 from location popularity, and 2 from the local coordinate system. 
\color{black}
We acknowledge that the data may vary across different regions due to differences in data completeness, user contributions, and regional coverage. However, by fusing multi-source data, we can mitigate the potential limitations of relying solely on a single data source to a certain extent, ensuring more reliable and representative POI profiles for each location. 
\color{black}
The location feature vector effectively captures geographical characteristics and facilitates trajectory generation by enabling location selection based on both movement patterns and location functions.

\textbf{Ethical Statement and Data Anonymization:} All input datasets used in this study, including gridded population distributions, Point of Interest (POI) maps, and Origin-Destination (OD) flows, were obtained from publicly available, aggregated, and fully anonymized sources. No individual-level data or personally identifiable information (PII) were accessed or processed. Furthermore, the resulting WorldMove dataset is purely synthetic, generated by a deep learning model to represent collective mobility patterns rather than individual behaviors. Consequently, this study does not involve human subjects research and was exempt from Institutional Review Board (IRB) approval.

\subsection*{Generating mobility trajectories via diffusion models}

Based on the collected location feature data, we employ a diffusion-based model~\cite{zhang2024noise} to generate mobility trajectories.
\color{black}
The core of our methodology is to resolve source-specific biases by learning mobility patterns in a universal, semantic space rather than from raw, heterogeneous geographic data. 
\color{black}
The whole process consists of three key steps. First, the multi-source location feature data is processed through a location feature encoder, compressing and projecting the regional characteristics into a unified embedding space.
\color{black}
This step creates a common language to describe all urban locations, regardless of their city of origin or the available data sources.
\color{black}
Building upon the location embeddings, we leverage real-world human mobility data to encode physical location sequences from different cities into a unified semantic space, forming a comprehensive mobility dataset that encompasses diverse urban mobility patterns. Our diffusion model is then trained on this unified dataset.
During the generation process, the diffusion model first generates a transition sequence within the embedding space. This embedding sequence is subsequently matched to the target city's location embeddings using a minimum-distance mapping, ultimately constructing the final mobility trajectory. 
The number of synthesized trajectories is, by default, proportional to the city’s total population from WorldPop to ensure representative urban scaling. Users can also specify custom sampling scales, which the framework then distributes according to spatial population density.
The detailed process is introduced in the following subsections.

\subsubsection*{Location encoding}
As described in Section~\ref{sec:2.2}, each location's feature is represented as a 38-dimensional vector $\mathbf{l}_i$, including attributes such as population $\mathbf{l}_i^{pop}$, POI distribution $\mathbf{l}_i^{poi}$, popularity rank $\mathbf{l}_i^{rank}$, and local coordinates $\mathbf{l}_i^{coord}$. To enable the generative model to comprehend regional characteristics across different cities, we employ an autoencoder~\cite{bank2021autoencoders} to learn these location features. This process compresses the feature vectors, reducing their dimensionality while projecting them into a unified embedding space that is more interpretable for the model. This transformation ensures that the model can effectively capture and generalize location attributes across diverse urban environments. The autoencoder is formulated as follows:

\begin{equation}
\label{eq:ae}
\begin{array}{l}
\mathbf{z}_i = \text{Encoder}(\mathbf{l}_i),\\
\widetilde{\mathbf{l}}_i = \text{Decoder}(\mathbf{z}_i).
\end{array}
\end{equation}

\noindent The autoencoder reduces feature dimensionality while preserving patterns by reconstructing location features. In our experiment, we set the embedding dimension to 8 and train a unified autoencoder on regions from six cities across China, the United States, and Africa. 
\color{black}
To ensure that the learned location embeddings capture the heterogeneous characteristics of different input features, we apply feature-specific loss functions during autoencoder training. This design reflects the semantic and statistical differences among feature types.  Specifically, we use MSE for continuous features (e.g., population density, spatial coordinates), KL divergence for POI distributions (treated as probability-like vectors), and cross-entropy for categorical features such as location popularity ranks.  This multi-objective loss formulation allows each type of information to be reconstructed under an appropriate metric, resulting in a more semantically meaningful embedding space. 
\color{black}

\begin{equation}
\label{eq:ae-loss}
\begin{array}{l}
\mathcal{L}(\mathbf{l}_i,\widetilde{\mathbf{l}}_i) = \|\mathbf{l}_i^{pop} - \widetilde{\mathbf{l}}_i^{pop}\|_2^2 + \|\mathbf{l}_i^{coord} - \widetilde{\mathbf{l}}_i^{coord}\|_2^2 + D_{kl}(\mathbf{l}_i^{poi},\widetilde{\mathbf{l}}_i^{poi}) + CE(\mathbf{l}_i^{rank},\widetilde{\mathbf{l}}_i^{rank}).
\end{array}
\end{equation}

\subsubsection*{Embedding-space generation}
To enable unified trajectory generation across cities, we first encode location features into a shared embedding space $\mathcal{Z}$ using a trained encoder. Each individual trajectory—originally a sequence of spatiotemporal points $\boldsymbol{p} = \{(l_1, t_1), (l_2, t_2), ..., (l_n, t_n)\}$—is converted into a sequence of embedding vectors $\mathbf{Z} = \{\mathbf{z}_1, \mathbf{z}_2, ..., \mathbf{z}_n\}$ with equal temporal intervals. 
\color{black}Here, $t$ is encoded using two temporal features: time of day and day of the week. The time of day captures the specific hour (within a 24-hour range), and the day of the week encodes whether the trajectory corresponds to a weekday or weekend (ranging from 0 to 6, representing Monday to Sunday).
\color{black}
These sequences are then reshaped into fixed-length vectors $\mathbf{x} \in \mathbb{R}^{48 \times 8}$ on a daily basis, forming the training set $\mathcal{X} = \{\mathbf{x}^1, \mathbf{x}^2, ..., \mathbf{x}^N\}$ aggregated from multiple cities.

We train a diffusion model $D_\theta$ over the embedding space to learn the generative distribution of human mobility. Following standard denoising diffusion probabilistic modeling, we learn a denoising network $\epsilon_\theta$ that reconstructs clean samples from noisy inputs:

\begin{equation}
\label{eq:diff-train}
\mathcal{L}(\theta) = \mathbb{E}_{\mathbf{x} \sim \mathcal{X}} \mathbb{E}_{\sigma \sim q(\sigma)} \mathbb{E}_{\mathbf{\epsilon} \sim \mathcal{N}\left(\mathbf{0}, \sigma^{2} \mathbf{I}\right)}\left\|D_{\mathbf{\theta}}(\mathbf{x}+\mathbf{\epsilon} ; \sigma)-\mathbf{x}\right\|_{2}^{2},
\end{equation}

\noindent where \( q(\sigma) \) defines the noise schedule. 
After training, the model generates embedding-space trajectories \( \mathbf{y} \) by denoising Gaussian noise through a deterministic DDIM sampling process~\cite{song2020ddim}. \color{black}
As part of the generation process, we incorporate collaborative noise priors derived from the city’s origin-destination (OD) data.

\begin{align} 
\label{eq:ddim} 
&\bm{z}_p = \text{Collab}_n(P, \bm{z}),\\
&\mathbf{y} = \text{DDIM}_\theta(\bm{z}), \quad \text{where } \bm{z} \sim \mathcal{N}(\bm{0}, \bm{I}).
\end{align}

\noindent Here, the noise prior \( \bm{z}_p \) is informed by the population OD flow $P$, providing additional guidance during the trajectory generation process. The details of how this collaborative noise mechanism  are described in CoDiffMob~\cite{zhang2024noise}.
\color{black}

\subsubsection*{Embedding-to-location decoding}  
After generating synthetic embedding sequences using the diffusion model, we decode them into real-world mobility trajectories by mapping each embedding back to a geographic location. Specifically, the diffusion model $D_{\theta}$ produces a sequence of location embeddings $\mathbf{y} = \{\widetilde{\mathbf{z}}_1, \widetilde{\mathbf{z}}_2, ..., \widetilde{\mathbf{z}}_n\}$, where each $\widetilde{\mathbf{z}}_i$ encodes latent spatial semantics.
To recover trajectories grounded in real geography, we employ a minimum-distance matching approach. First, we use the encoder to compute the set of location embeddings $L_{\mathbf{z}} = \{\mathbf{z}_1, \mathbf{z}_2, ..., \mathbf{z}_m\}$ from the geographic locations in the target city. Then, each generated embedding $\widetilde{\mathbf{z}}_i$ is mapped to its nearest neighbor in $L_{\mathbf{z}}$ by:

\begin{equation}
\label{eq:loc-proj}
\text{index} = \mathop{\arg\min}_j \|\mathbf{z}_j - \widetilde{\mathbf{z}}_i\|_2^2,
\end{equation}

\noindent which yields a location sequence $\{l_1, l_2, ..., l_n\}$ corresponding to real-world coordinates. Finally, we assign timestamps $\{t_1, t_2, ..., t_n\}$ based on fixed time intervals to form a complete spatiotemporal trajectory $\{(l_1, t_1), (l_2, t_2), ..., (l_n, t_n)\}$.
This decoding process enables the model to generate real-world mobility trajectories for any given city, by conditioning on city-specific embedding spaces while maintaining semantic consistency across regions.

\subsection*{Ethics statement.}

In this section, we outline the measures implemented to safeguard user privacy and eliminate potential risks associated with utilizing mobility trajectory data.
First, the Terms of Service for mobile operations explicitly include user consent for research purposes, ensuring ethical data usage.
Second, all potential individual identifiers are replaced with anonymized hash codes, preventing any records from being linked to specific individuals.
Third, all research data is securely stored on offline servers with access strictly limited to authorized researchers under binding confidentiality agreements. These measures collectively ensure robust data security and compliance with ethical standards.

\section*{Data Records}

\subsection*{Provided datasets}
The released dataset includes synthetic mobility data for over 1,600 cities across 179 countries and 6 continents, covering a diverse range of urban environments. To support open research and encourage broad use, we provide access to the dataset through multiple platforms.
A static version is archived on Figshare~\cite{worldmovefigshare2024} to ensure long-term accessibility and citation. Meanwhile, a live version is maintained and regularly updated on our GitHub repository\footnote{\url{https://github.com/tsinghua-fib-lab/WorldMove}}. In addition, we offer an interactive web portal\footnote{\url{https://fi.ee.tsinghua.edu.cn/worldmove}}, where users can explore cities on a global map and directly download data for any selected city. Alternatively, users can search by city name to quickly locate and retrieve the corresponding dataset.
For each city, we organize the data into a dedicated folder, which includes synthetic individual mobility trajectories, city boundary shapefiles, regional divisions, POI distributions, and commuting OD matrices. The folder structure is designed to be intuitive and easy to navigate, enabling seamless integration into downstream research tasks. 

\subsubsection*{City boundary and location division}  The city boundary data is provided in shapefile format, ensuring compatibility with various geographic information systems. Additionally, location data is indexed, with each index corresponding to the longitude and latitude of the center of its respective grid.

\subsubsection*{Mobility trajectory data} 
\begin{figure}[p]
\centering
\includegraphics[width=\linewidth]{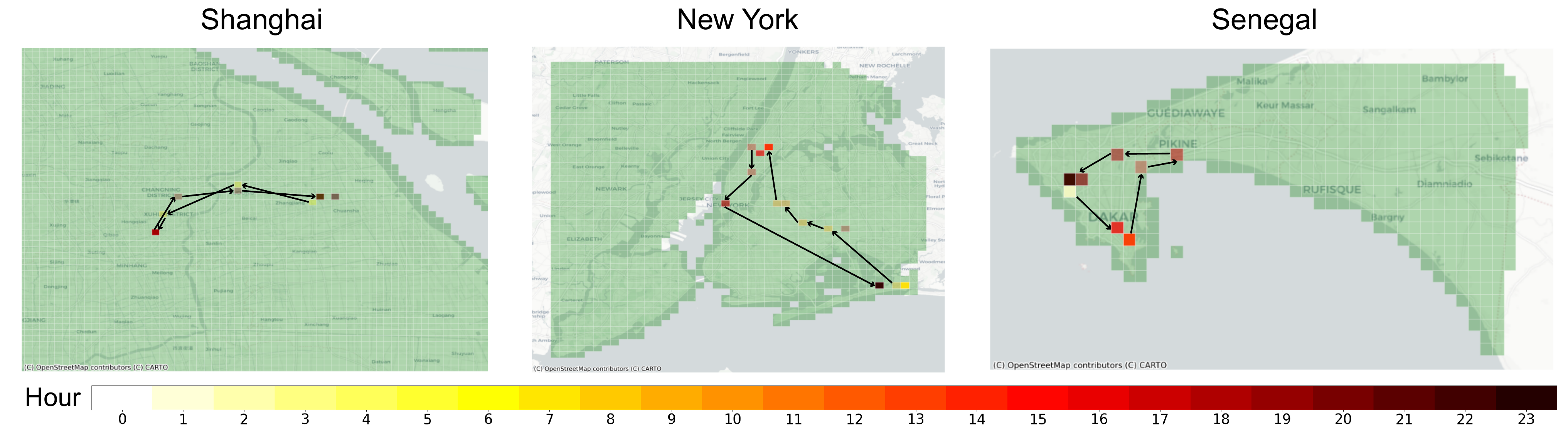}
\caption{Example of the generated trajectories for three cities.}
\label{fig:traj}
\end{figure}

\begin{figure}[p]
\centering
\includegraphics[width=0.8\linewidth]{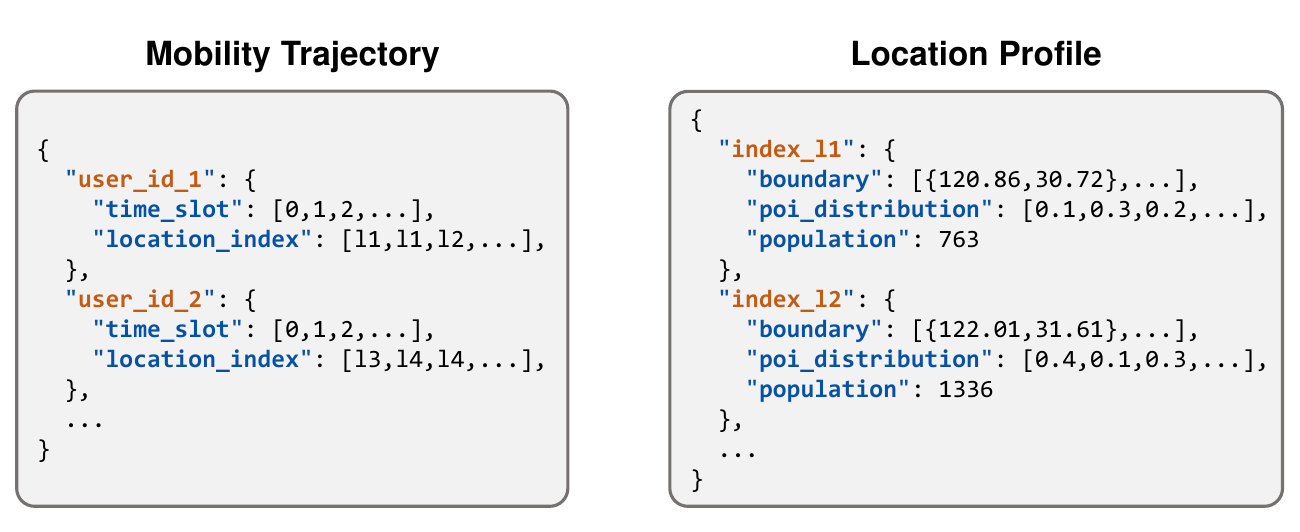}
\caption{Examples of the mobility trajectory dataframe and the corresponding location profile dataframe. 
}
\label{fig:data_example}
\end{figure}

Figure~\ref{fig:traj} illustrates examples of the generated trajectories for three cities. We provide the mobility trajectory data in a dictionary (Dict) format (see Figure~\ref{fig:data_example}), which contains multiple user trajectories. In this structure:

\begin{itemize}[leftmargin=*]
    \item \textbf{Key:} Each key represents a unique User ID, which is an anonymized identifier assigned to individual users. This ensures privacy while allowing analysis at the individual level.
    \item \textbf{Value:} The value associated with each User ID is a sequence of tokens. Each token contains:
    \begin{itemize}
        \item \textbf{Time slot:} A half-hour time interval represented by an index. For example, 0 corresponds to 00:00 AM–00:30 AM, and 27 corresponds to 1:30 PM–2:00 PM.
        \item \textbf{Location index:} The grid index represents a specific location within the city’s spatial grid. Each index corresponds to the center of a grid cell, ensuring precise spatial representation.
    \end{itemize}

\end{itemize}

\subsubsection*{Location profiling}  
The location profiling data is provided in a dictionary (Dict) format. Each key represents a \textit{location index}, and the corresponding value is another dictionary. This nested dictionary contains the following keys:

\begin{itemize}[leftmargin=*]
    \item \textbf{"POIs"}: Represents the profile of points of interest for the location, detailing the number and types of POIs within the grid.
    \item \textbf{"Popularity"}: Indicates the location's popularity, typically represented by a visitation frequency or rank derived from mobility patterns.
    \item \textbf{"Population"}: Provides the population density or estimate for the location.
\end{itemize}

\noindent This structure facilitates a comprehensive representation of each location by encapsulating key attributes within an easily accessible format.

\subsection*{Mobility Data Generation Pipeline}
In addition to the provided datasets, we also offer a pre-trained model and a pipeline tool for generating mobility datasets for any city worldwide. The pipeline follows a straightforward process:  

\begin{itemize}[leftmargin=*]
    \item \textbf{Step 1: Acquire population data.} The population data can be obtained from WorldPop using the script \texttt{Generate\_pop.py}.  
    \item \color{black} \textbf{Step 2: Generate mobility data.} Using the prepared population data, mobility data can be generated with the script \texttt{generate\_mobility.py}, which now integrates the process of creating location profiles based on population distribution and POI attributes. \color{black}
\end{itemize}

\section*{Technical Validation}

In this section, we evaluate the quality of the generated mobility data to demonstrate that it not only adheres to the characteristics and patterns observed in real-world data but can also serve as a reliable substitute for real-world data in downstream applications. The technical validation is conducted across four key aspects: (1) data fidelity, by comparing the generated data with real-world data; (2) reproduction of mobility laws, by examining key distributions of mobility characteristics; (3) alignment with commuting origin-destination (OD) flows, which represent aggregated mobility patterns; and (4) privacy evaluation, to ensure that the generated data safeguards user privacy by preventing the retrieval of individual-level information.

\color{black}

\subsection*{Real-world data}

To train and validate the diffusion-based generative model, we utilize real-world human mobility trajectory datasets collected from six diverse cities across three continents: Shanghai, Nanchang (China), and Senegal (Africa) via mobile operator and CDR data; New York, Los Angeles, and Chicago (North America) via anonymized credit card transaction records. These datasets collectively capture a wide range of urban forms, socioeconomic structures, and mobility behaviors, thus enhancing the model's generalizability across global contexts. Table~\ref{tab:realdata} summarizes the key statistics of the real-world trajectory data used in this study, including data source, spatial scope, temporal duration, number of users, and trajectory size. Importantly, we include both data-rich cities (e.g., Shanghai, New York) and data-scarce regions (e.g., Nanchang, Senegal) to evaluate the model’s generalizability under heterogeneous data conditions. This directly supports our motivation to develop a globally transferable model, particularly for low-resource regions.

These real-world datasets are used to (1) train the location embedding autoencoder on a semantically rich representation space, (2) train the diffusion model over unified trajectory embeddings, and (3) quantitatively validate the generated synthetic data in terms of mobility metrics, OD flows, and privacy risks. While the datasets cannot be released due to licensing constraints, they enable reliable benchmarking of our model’s performance under varied geographic and behavioral contexts.

\color{black}

\subsection*{Data fidelity}

We evaluate data fidelity from two perspectives: individual trajectory fidelity and population flow consistency. 
At the individual level, we assess the statistical alignment of the generated trajectories with real-world data using distribution differences, measured by Jensen-Shannon divergence (JSD) and the Kolmogorov-Smirnov (KS) test~\cite{jiang2016timegeo,feng2020learning}. Both metrics range from $[0,1]$, with 0 indicating perfect match. We evaluate several key mobility metrics~\cite{barbosa2018human,wang2019urban,luca2021survey}, including jump length ($\Delta r$),  daily trip distance ($r_w$), radius of gyration ($r_g$), waiting time ($\Delta t$), and daily visited locations ($S_d$).Additionally, we assess adherence to Zipf's law~\cite{zipf1949human}, which captures the power-law distribution of location visit frequencies in real-world mobility.
These metrics assess spatial and temporal regularities ($\Delta r$, $\Delta t$)~\cite{gonzalez2008understanding}, individual variability ($r_g$)~\cite{gonzalez2008understanding}, and the slow growth of travel distances and visited locations ($r_w$, $S_d$)~\cite{song2010modelling}.
We also evaluate aggregated mobility patterns (population flows), comparing observed and generated flows by calculating metrics such as Root Mean Square Error (RMSE) and common parts of commuting (CPC). CPC ranges from [0,1], with 1 indicating perfect correlation, while RMSE quantifies the difference between the distributions, further validating data accuracy. These two metrics are defined as follows:

\begin{align}
    &\text{JSD}(d_1, d_2) = \frac{1}{2}D_{KL}(d_1\|m) + \frac{1}{2}D_{KL}(d_1\|m), \quad \text{where } m = \frac{1}{2}(d_1 + d_2), \nonumber \\
    &\text{KS}(d1, d2) = \text{sup}_x \|F_{d_1}(x) -  F_{d_2}(x)\|,\quad \text{where }F_d(x)=\frac{1}{n}\sum_{i=1}^n\mathbb{I}_{(-\inf,x]}(X_i)\text{ and }X_i\sim d(x), \nonumber \\
    &\text{RMSE} = \sqrt{\frac{1}{N} \sum_{i=1}^{N} \sum_{j=1}^{N} (f_{ij}^{\text{obs}} - f_{ij}^{\text{gen}})^2} ,\nonumber \\
    &\text{CPC} =  \frac{\sum_{i=1}^N\sum_{j=1}^N\min(f_{ij}^{\text{obs}},f_{ij}^{\text{gen}})}{\sum_{i=1}^N\sum_{j=1}^N f_{ij}^{\text{obs}} + \sum_{i=1}^N\sum_{j=1}^N f_{ij}^{\text{gen}}} ,\nonumber
\end{align}

\noindent where \( f_{ij}^{\text{obs}} \) Represents the observed flow for location pair \( (i,j) \), \( f_{ij}^{\text{gen}} \) Represents the generated flow for location pair \( (i,j) \), and \( N \): The total number of location pairs.

Through our collaboration with mobile operators and credit card companies, we have access to mobility trajectory data from multiple cities across China, Senegal, and the US. This enables a comprehensive evaluation of the quality of the generated data. While we are unable to release real-world data due to strict non-disclosure agreements (NDAs), we leverage them to assess the performance of our generated data. 
\color{black}Table~\ref{table:model_perf} presents the numerical results under a zero-shot transfer setting, where the model is trained without access to target data.\color{black}  The table includes individual-level metrics (JSD and KS test) and population-level metrics (RMSE and CPC). 
The generated data achieves JSD and KS values below 0.05 and 0.28, respectively, indicating strong alignment with the real-world distribution of mobility characteristics. These results represent state-of-the-art performance in the field of mobility trajectory generation, surpassing prior benchmarks where JSD and KS values typically range around 0.07~\cite{feng2020learning, chu2024simulating,long2023practical} and 0.30~\cite{jiang2016timegeo}. 
 In terms of CPC, our method reaches a value of 0.41, significantly higher than the 0.32 reported in recent studies~\cite{li2023learning}, further confirming that population-level movement patterns in our synthetic data are highly consistent with real-world behavior.
 \color{black}
We also show the results where some target data is used to fine-tune the model, as shown in Table~\ref{table:model_perf_insample}. The performance improves significantly, implying that performance can be enhanced when some data is available for the target city.
 \color{black}
To present the results in more detail, we plot the distribution of these key metrics in Figure~\ref{fig:law} and Figure~\ref{fig:law_usa}, which clearly demonstrate that the generated data closely matches the real-world data.

\subsection*{Reproduction of mobility laws}

\begin{figure}[p]
\centering
\includegraphics[width=\linewidth]{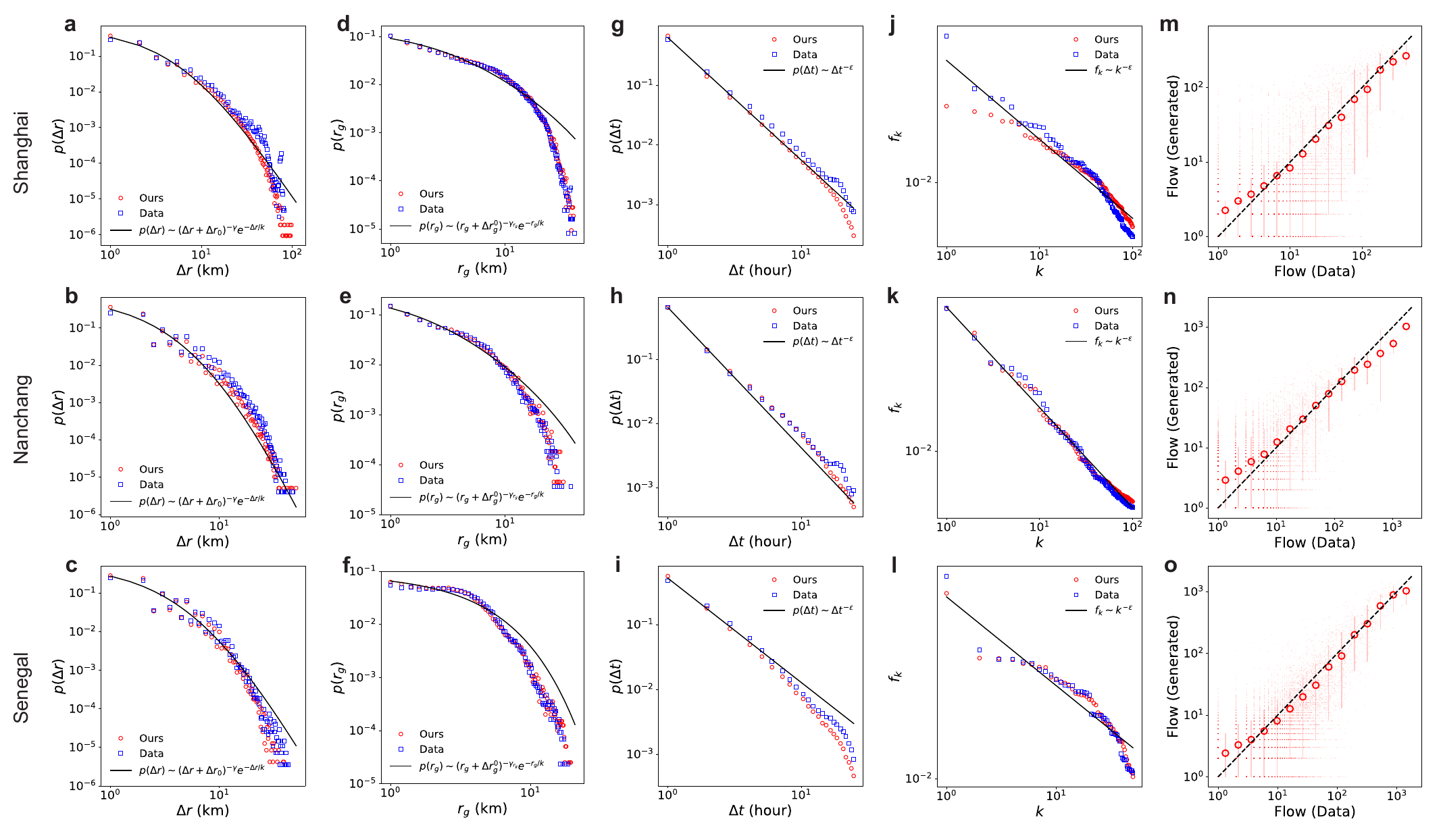}
\caption{Evaluation of key mobility pattern reproduction across three cities: Shanghai, Nanchang, and Senegal.  (a–c) Jump length; (d–f) radius of gyration; (g–i) waiting time; (j–l) rank-frequency relationship following Zipf’s law; (m–o) comparison between model-predicted and real population flows.}
\label{fig:law}
\end{figure}

\begin{figure}[p]
\centering
\includegraphics[width=\linewidth]{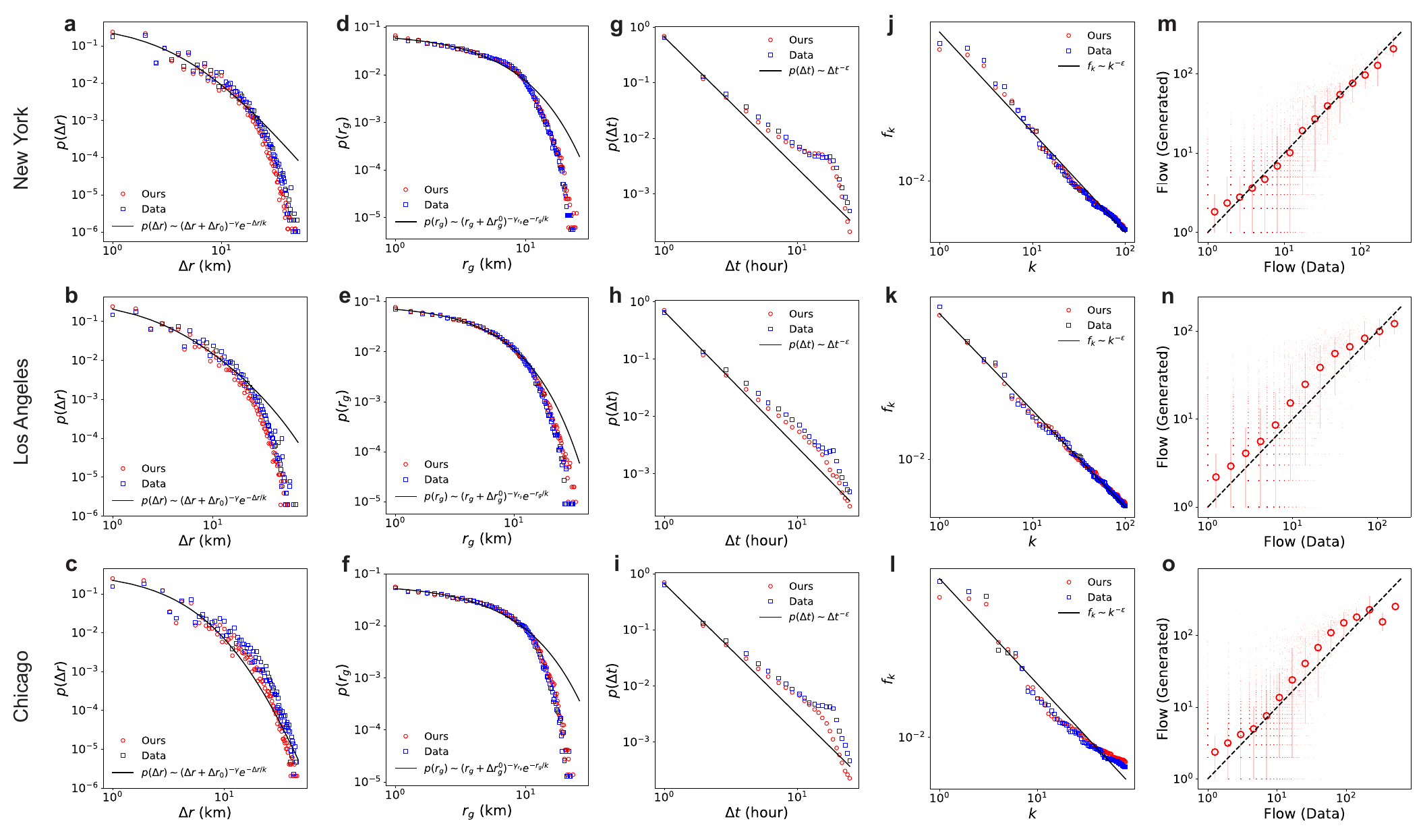}
\caption{Evaluation of key mobility pattern reproduction across three cities: New York, Los Angeles, and Chicago.  (a–c) Jump length; (d–f) radius of gyration; (g–i) waiting time; (j–l) rank-frequency relationship following Zipf’s law; (m–o) comparison between model-predicted and real population flows.}
\label{fig:law_usa}
\end{figure}

Human mobility has been extensively studied, leading to the identification of several mobility laws documented in the literature~\cite{}. To evaluate whether the generated data not only resemble real-world data but also adhere to fundamental mobility laws, we analyze the generated mobility patterns and visualize them in Figure~\ref{fig:law} and Figure~\ref{fig:law_usa}.
The jump length $\Delta r$, which measures the spatial distance between consecutive stops, follows a truncated power-law distribution: $p(\Delta r) \sim (\Delta r+\Delta r_0)^{-\gamma_1}\text{exp}(-\Delta r/\kappa_1)$, with the scaling exponent $\gamma_1$$\sim (0.92-1.39$),  which remains consistent across different cities and aligns closely with empirical values~($\sim 1.1-1.3$)~\cite{gonzalez2008understanding}.  Similarly, the radius of gyration, which quantifies the spatial spread of an individual’s movements, also follows a truncated power-law distribution, with $\gamma_2\sim(1.29, 1.71)$~\cite{gonzalez2008understanding,song2010modelling}.   The distribution of waiting times, $\Delta t$, follows a power-law form $p(\Delta t)\sim \Delta t^{-\epsilon}$ with $\epsilon \sim (2.04-2.37)$, further reflecting the mobility laws observed in real-world data.
In addition to these, we explore another critical scaling property of human mobility: Zipf's law~\cite{gonzalez2008understanding}, which characterizes the frequency distribution of visited locations.  As shown in Figure~\ref{fig:law} and Figure~\ref{fig:law_usa}, it follows a power-law form: $f_k \sim k^{-\zeta}$($\zeta \sim 0.42-0.61$), closely matching the empirical values~\cite{gonzalez2008understanding}. 
These results demonstrate that the generated data do not merely fit real-world data, but rather reflect the ability of our generation method to capture the fundamental mobility laws that govern human movement.

\color{black}

\subsection*{Visualization of embedding distributions}

\begin{figure}[p]
\centering
\includegraphics[width=1.0\linewidth]{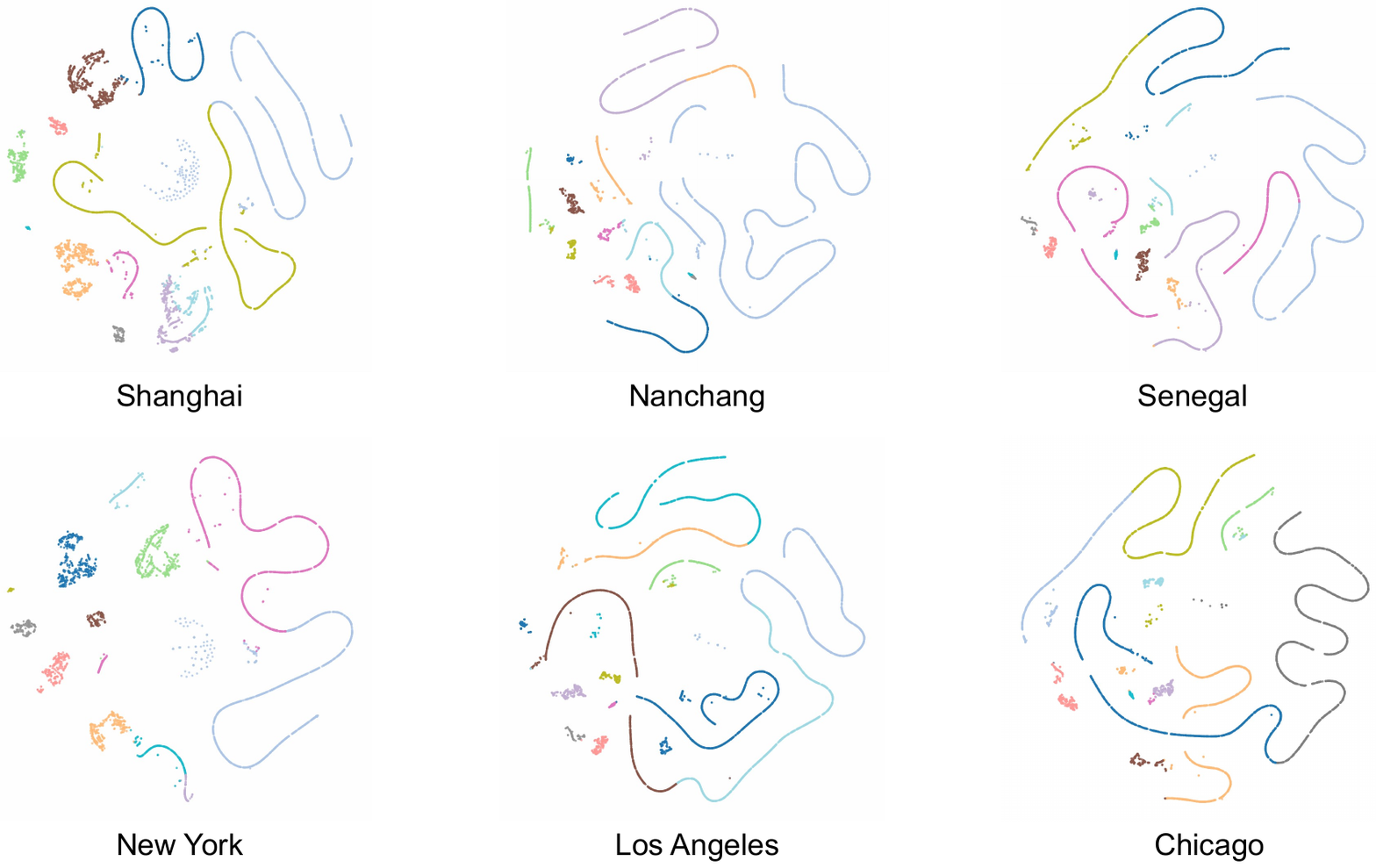}
\caption{T-SNE visualization of location embeddings across six cities. Each point represents a grid-level location embedding learned by the autoencoder, and colors indicate clusters obtained via K-Means. Densely clustered points typically correspond to high-activity areas with rich POI distributions and frequent visits, while elongated or dispersed regions reflect low-activity or underdeveloped zones. The visualization reveals both shared structural patterns and city-specific distinctions in urban morphology and functional layout.}
\label{fig:region_embed}
\end{figure}

To directly assess the outcome of our multi-source harmonization methodology, we visualize the location embeddings obtained from the trained autoencoder. This analysis serves to verify that the model successfully learns a unified and semantically meaningful representation space from the diverse training data. 
Specifically, we apply T-SNE for dimensionality reduction and K-Means clustering to identify latent functional groupings among city locations. Figure~\ref{fig:region_embed} presents the resulting embedding distributions for six cities: Shanghai, Nanchang, Senegal, New York, Los Angeles, and Chicago. Each point corresponds to a 1km × 1km grid cell in the city, projected into the 2D space from the learned 8-dimensional embedding. The colors indicate K-Means cluster assignments. From the visualization, we observe two key findings:

\begin{itemize}[leftmargin=*]
    \item \textbf{Semantic regularity across cities:} Across all cities, clusters representing high-population, high-POI regions (e.g., central business districts, major residential zones) form dense and compact groups, demonstrating that the model captures shared semantic structure across urban environments. These clusters typically correspond to highly visited or functionally active areas.
    \item \textbf{City-specific structural heterogeneity:} While certain mobility patterns are consistent across cities, each city also exhibits distinct patterns in the shape, density, and dispersion of its clusters. For example, Los Angeles and Chicago show more fragmented embeddings, potentially reflecting polycentric urban layouts, while cities like Shanghai and New York exhibit stronger centralization in their cluster distributions. This suggests the model also retains meaningful distinctions between cities, aligning with their respective urban morphologies.
\end{itemize}

Overall, this embedding analysis demonstrates that the model learns a semantically coherent yet city-sensitive representation space, which supports both the generalization to unseen cities and the preservation of local urban semantics. These properties are essential for reliable cross-region mobility generation and further confirm the model’s robustness and transferability.

\color{black}

\subsection*{Visualization of commuting OD flows}
\begin{figure}[p]
\centering
\includegraphics[width=\linewidth]{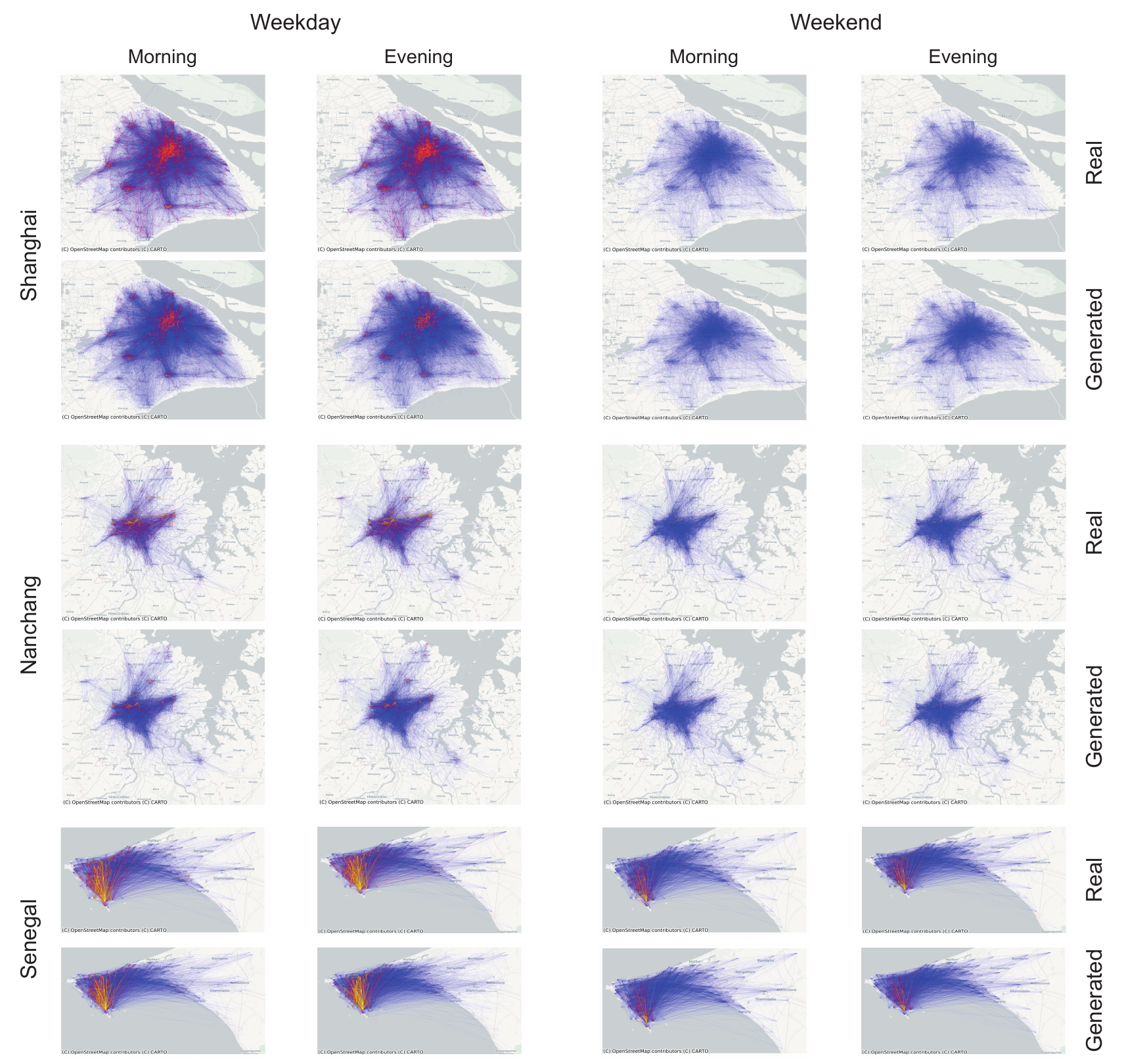}
\caption{Visualization of aggregate mobility patterns in Shanghai, Nanchang, and Senegal, illustrating OD flows during morning and evening peak hours on both weekdays and weekends.
\color{black} Blue edges indicate flows with a number of commuters between 0 and 5, red edge between 5 and 15, and yellow edges above 15 commuters. \color{black}
}
\label{fig:viz}
\end{figure}

\begin{figure}[p]
\centering
\includegraphics[width=\linewidth]{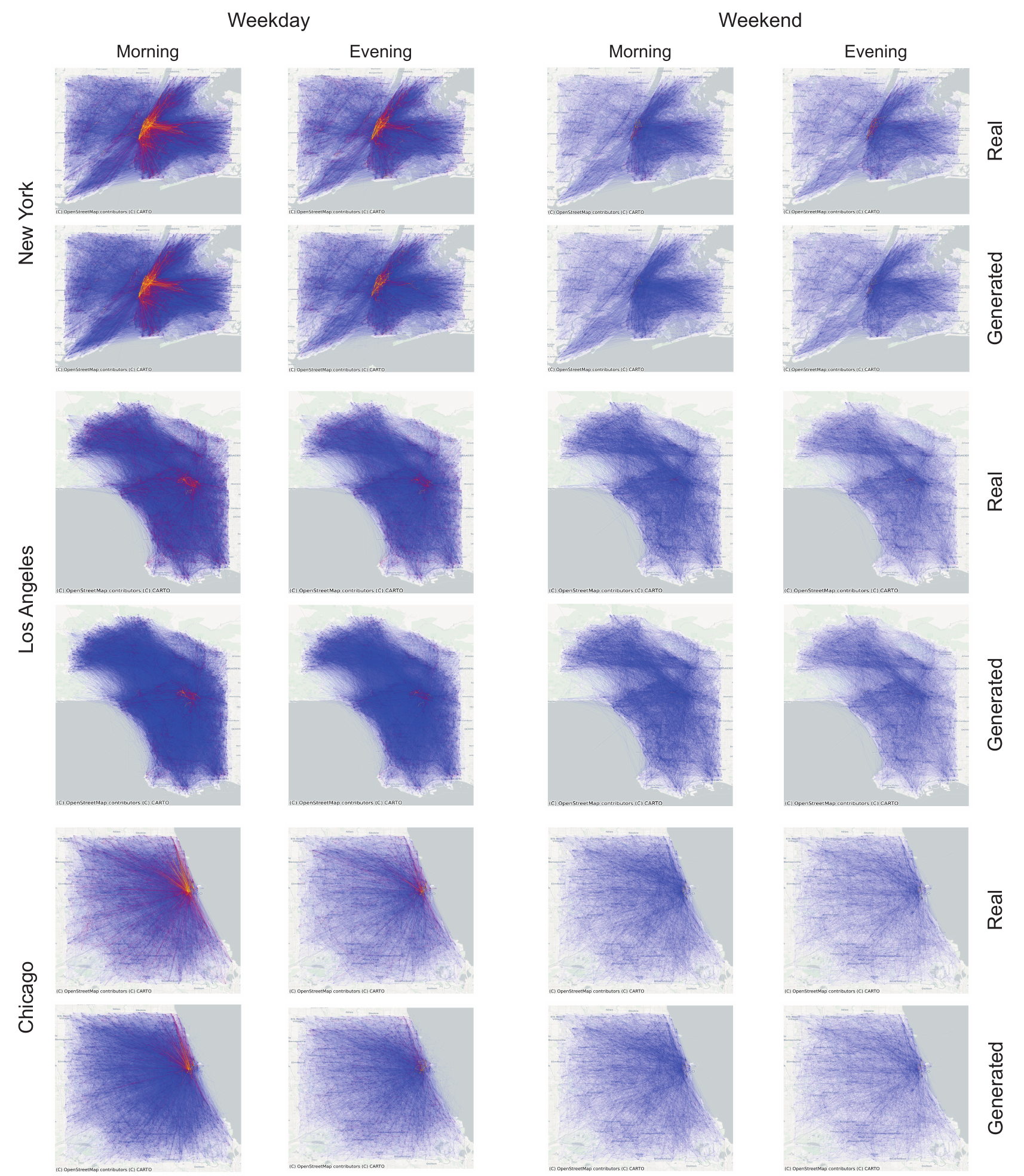}
\caption{Visualization of aggregate mobility patterns in New York, Los Angeles, and Chicago, illustrating OD flows during morning and evening peak hours on both weekdays and weekends. \color{black} Blue edges indicate flows with a number of commuters between 0 and 2, red edges
 between 2 and 5, and yellow edges above 5 commuters. \color{black}}
\label{fig:viz_usa}
\end{figure}

While WorldMove generates individual-level mobility trajectories, it is important to ensure that these fine-grained movements can accurately reproduce aggregated urban mobility patterns. One key proxy for such patterns is the commuting origin-destination (OD) flow~\cite{rong2024interdisciplinary}, which reflects where people travel from and to during typical commuting hours. 
To intuitively demonstrate that the generated trajectories accurately reflect commuting OD flows, we visualize the distribution of these flows during morning and evening commuting times in Figure~\ref{fig:viz} and Figure~\ref{fig:viz_usa}. We include cities from various continents, including China, Africa, and the Americas, to provide a diverse range of examples.  

As we can observe, the generated OD flows closely resemble the real-world patterns, indicating that the generated mobility data can capture aggregated mobility trends effectively.
In addition, the generated data reflects regional heterogeneity in commuting structures. 
For example, in large metropolitan areas such as Shanghai and several U.S. cities, we observe clear directional flows from peripheral residential zones to central business districts in the morning, and the reverse in the evening—patterns typical of centralized urban employment. On weekends, this structured pattern becomes less prominent, consistent with reduced work-related travel. In contrast, in medium-sized cities such as Nanchang, China, where residential and workplace zones are less spatially segregated, the OD flows are more evenly distributed. The generated data accurately captures these distinctions, further validating its ability to reproduce nuanced urban commuting behaviors.

\color{black}

\subsection*{Visualization of visitation heatmaps}

\begin{figure}[p]
\centering
\includegraphics[width=1.0\linewidth]{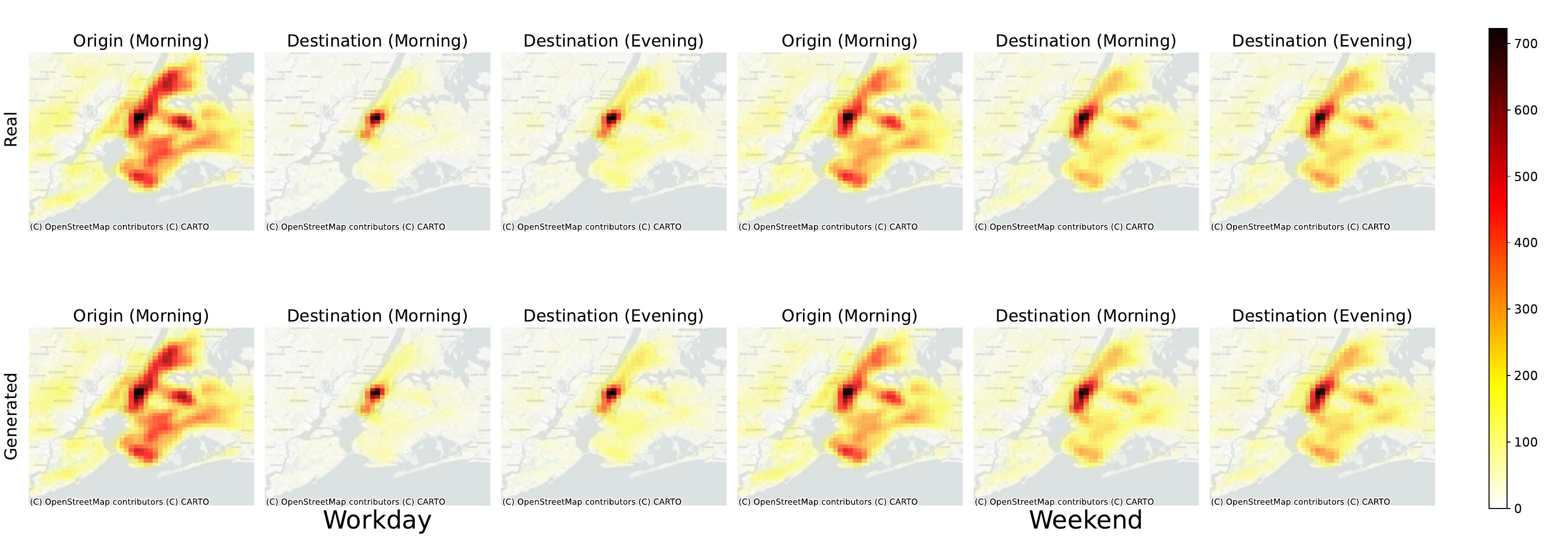}
\caption{Dynamic population density heatmaps of New York. The top row shows real-world population density during morning and evening commutes across workdays and weekends, while the bottom row presents the generated population density.}
\label{fig:pop_density}
\end{figure}

To validate the quality and realism of the synthetic data, we present comparative visualizations of visitation heatmaps derived from both real-world and generated trajectories. These heatmaps highlight key hotspots for origin and destination flows during morning and evening commutes, as well as across workdays and weekends. The top row of Figure~\ref{fig:pop_density} shows the real-world OD flows, while the bottom row presents the generated flows, allowing for a direct comparison of spatial and temporal dynamics.

In terms of morning versus evening rush hours, both the real and synthetic data show concentrated hotspots in central regions during the morning rush hour (typically between 7-9 AM), with origin flows from residential areas towards business districts. Similarly, in the evening rush hour (5-7 PM), destination flows shift towards residential areas. The synthetic data closely follows this spatial shift, with morning origin flows and evening destination flows aligning well with real-world patterns, though minor variations in intensity and geographical distribution can be observed. Regarding weekday and weekend mobility, both real and generated data exhibit heavy flows in central business districts during weekdays. However, on weekends, the mobility patterns become more dispersed, with a shift towards recreational areas, shopping centers, and parks, which the synthetic data captures effectively. Minor differences in flow density in suburban regions may occur, which could be attributed to data sparsity in these areas.

\color{black}

\color{black}

\subsection*{Distribution of frequent mobility networks}

\begin{figure}[p]
\centering
\includegraphics[width=1.0\linewidth]{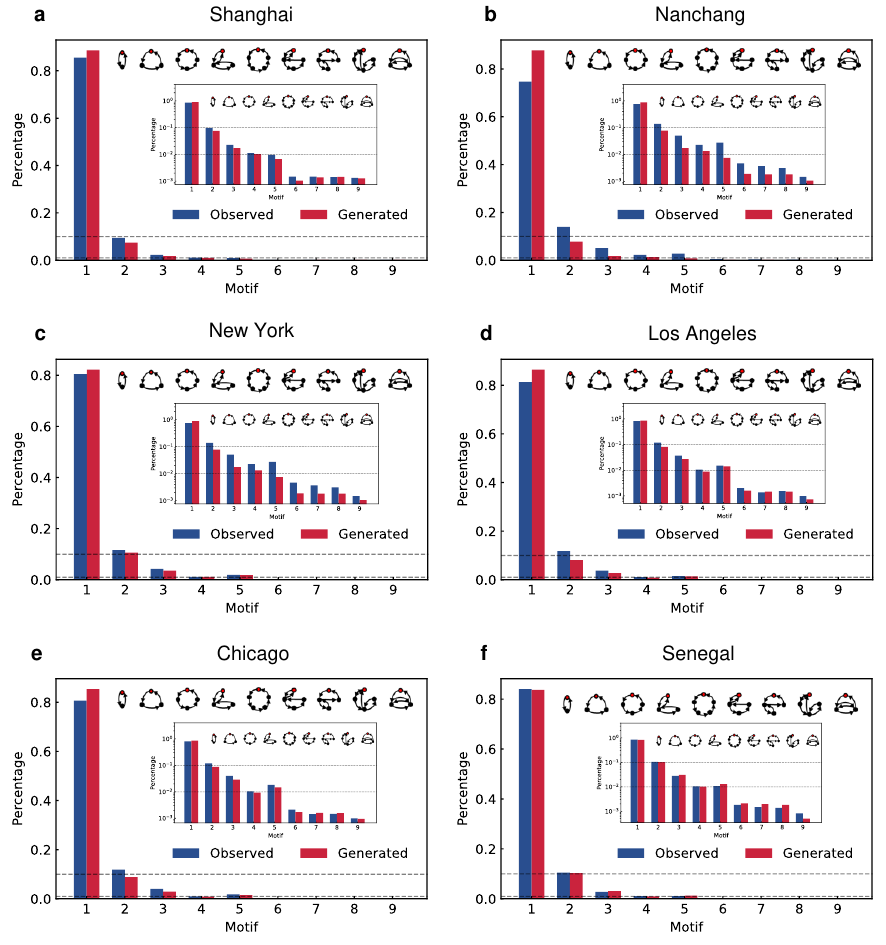}
\caption{Distribution of mobility motifs, which presents the motif distribution of real-world and generated trajectories across six cities. There is a strong alignment between the real and generated data, with a few dominant motifs (e.g., home-to-work routes) accounting for a significant portion of the distribution.
}
\label{fig:motif}
\end{figure}

To better understand the travel patterns and frequent paths in the synthetic dataset, we visualize the motif (mobility network) distribution of both real-world and generated trajectories in Figure~\ref{fig:motif}. The motif analysis reveals that the real trajectories and generated motifs are highly similar. We observe that a small number of motifs occupy a significant portion of the total distribution (e.g., common routes such as home-to-work), while other motifs appear less frequently. This pattern is consistent with findings from previous studies~\cite{jiang2016timegeo,schneider2013unravelling}, which indicate that a few key mobility patterns dominate most human mobility behavior. The alignment between real and synthetic motifs reinforces the idea that the generated data effectively captures the underlying mobility structure of urban environments.

\color{black}

\subsection*{Privacy protection evaluation}

As a synthetic global dataset, we evaluate whether the generated data protects individual privacy and avoids potential information leakage. To this end, we conduct a membership inference attack, which is a common method used to assess the risk of privacy breaches. The attack attempts to infer whether a particular data sample was part of the training set, which could indicate potential memorization or leakage of sensitive information.
In our evaluation, we train a binary classifier to distinguish between real (training) and synthetic samples based on their latent representations. 
To ensure the robustness of the results, we adopt three commonly used binary classifiers: logistic regression (LR), support vector machines (SVM), and random forests (RF).

\begin{figure}[p]
\centering
\includegraphics[width=\linewidth]{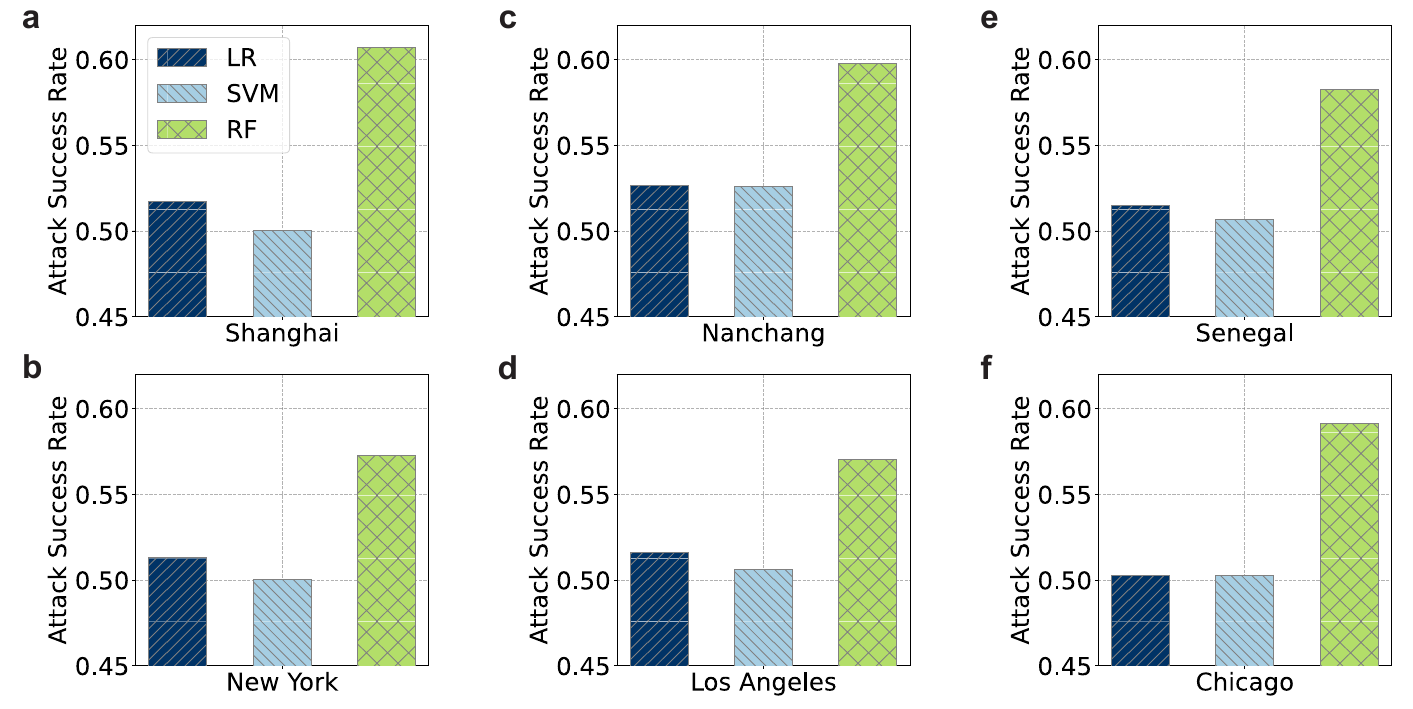}
\caption{Privacy evaluation using membership inference attacks with three representative binary classifiers.}
\label{fig:privacy}
\end{figure}

As shown in Figure~\ref{fig:privacy}, the attack success rate remains between 0.5 and 0.6 across different cities and scenarios, which is close to random guessing (0.5). This indicates that the diffusion model does not memorize specific individual trajectories and that the generated data does not reveal identifiable information about real users. These results confirm that our synthetic dataset offers strong privacy protection, making it suitable for open sharing and safe use in research and application scenarios involving human mobility data.

\color{black}

\subsection*{Performance under different model sizes}

\begin{figure}[p]
\centering
\includegraphics[width=0.6\linewidth]{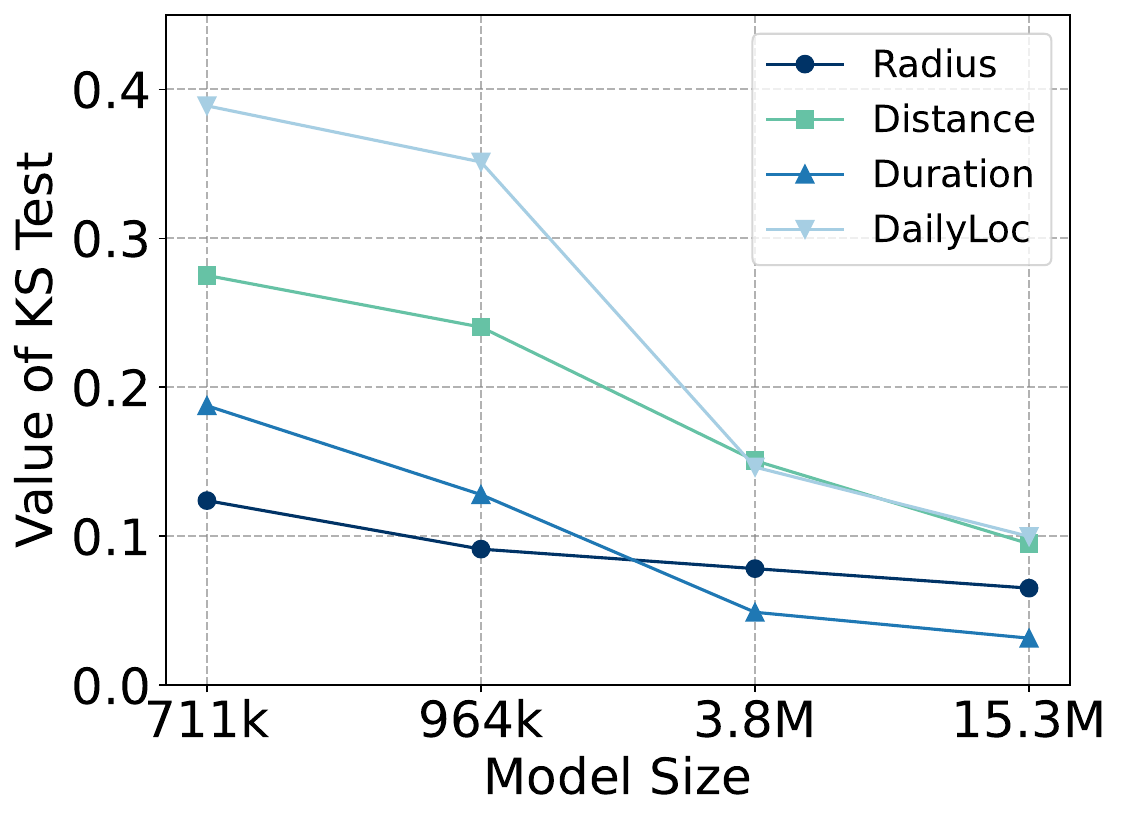}
\caption{Evaluation of model performance with varying model sizes (ranging from 711k to 15.3M parameters). Larger models improve alignment between generated and real-world trajectories, with diminishing returns observed beyond an optimal model size.}
\label{fig:scale}
\end{figure}

To evaluate the sensitivity of the pipeline to model parameters, we performed experiments using different model sizes, ranging from 711k to 15.3M parameters. The results in Figure~\ref{fig:scale} show that increasing the model size improves the alignment between generated and real-world trajectories. Specifically, larger models achieve lower KS test values across multiple metrics, indicating better reproduction of key mobility characteristics such as radius, distance, duration, and daily location visits. However, we also observe diminishing returns as the model size increases, suggesting that while larger models improve the quality of generated trajectories, there is an optimal scale beyond which additional increases in model size yield limited improvements.

\color{black}

\section*{Usage Notes}

\subsection*{Data and model usage across conditions}

\color{black}

Since WorldMove includes both the generated dataset, the trained model, and the complete pipeline for data generation, we offer two types of usage for the WorldMove project:

\begin{itemize}[leftmargin=*]
\item \textbf{Zero-shot generation for cities with no data:} For cities with no real-world data, our zero-shot approach is employed. While this leads to a slight performance reduction, it still generates useful synthetic data. This is particularly valuable for regions where data collection is challenging or impossible, enabling data-driven research and decision-making in data-scarce areas.

\item \textbf{Fine-tuning for cities with partial data:} For cities where some real-world data is available, the model can be fine-tuned to improve both individual and population-level accuracy. We find that even a small amount of real-world data greatly enhances the model’s ability to generate realistic individual trajectories and population flows. 
\end{itemize}

\noindent This adaptability enhances the practical utility of the WorldMove framework, ensuring it can be applied in diverse urban contexts, from well-connected, data-rich cities to remote and underserved regions, thereby confirming its practical utility for global mobility research.

\color{black}

\subsection*{WorldMove’s application in research}

\begin{figure}[p]
\centering
\includegraphics[width=\linewidth]{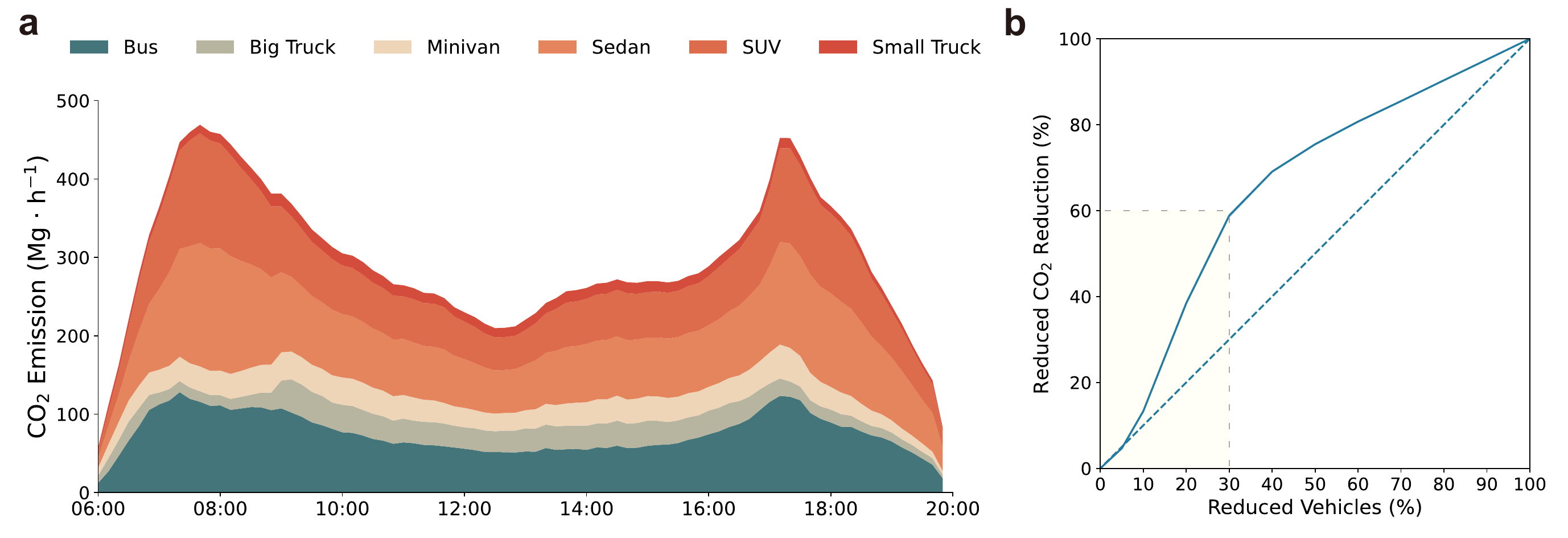}
\caption{Application of the generated trajectory data on traffic emission analysis.}
\label{fig:app_carbon}
\end{figure}

\begin{figure}[p]
\centering
\includegraphics[width=0.8\linewidth]{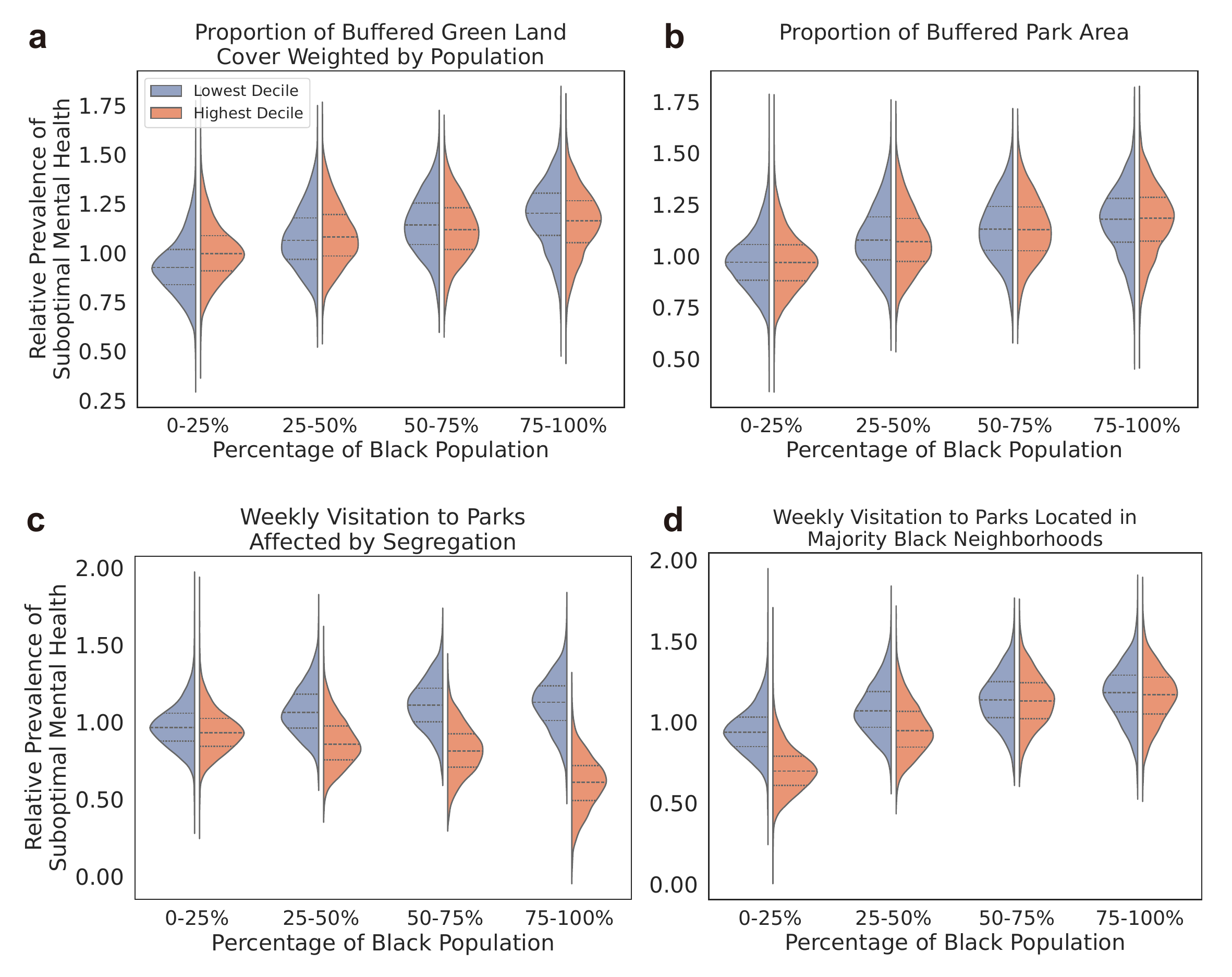}
\caption{Application of the generated trajectory data on inequity analysis on greenness segregation and mental health.}
\label{fig:app_greeness}
\end{figure}

The generated mobility data serves as a versatile and valuable resource for a wide range of research fields. By simulating realistic human mobility patterns, it can support the development and evaluation of models, algorithms, and policies in areas where real-world data may be limited or difficult to obtain.  Below are some key areas where this data can be applied:

\begin{itemize}[leftmargin=*]
    \item \textbf{Sustainable transportation optimization:}  WorldMove enables realistic simulation of urban transportation systems and carbon emissions analysis. In our case study, we integrate WorldMove-generated human mobility data with vehicle type information~\cite{Yu2023trajectory}, and simulate daily urban traffic using the MOSS microscopic traffic simulator~\cite{zhang2024mosslargescaleopenmicroscopic}. 
    We simulate carbon emissions with the transportation simulation~\cite{ehsani2016modeling} across six vehicle types over a day, which reveals clear peaks during commuting hours (Figure~\ref{fig:app_carbon}a). We further conduct a counterfactual experiment by reducing vehicle numbers during peak periods. As shown in Figure~\ref{fig:app_carbon}b, a 30\% reduction in vehicles results in a 60\% drop in emissions, highlighting the nonlinear benefits of congestion mitigation. This example illustrates WorldMove’s potential to support data-driven research on sustainable transportation and emission reduction strategies.

    \item \textbf{Urban environment and exposure analysis:}  The dataset supports research in environmental exposure modeling, such as estimating population exposure to green spaces, noise, or air pollution. As an example, we analyze the relationship between residents’ exposure to green spaces and mental health outcomes in multiple U.S. cities (See \url{https://github.com/tsinghua-fib-lab/MentalHealthInequity} for details). We combine WorldMove-generated mobility trajectories with land-use maps, community demographic profiles, and mental health index data. As shown in Figure~\ref{fig:app_greeness}, traditional static indicators (e.g., green space coverage) do not fully explain disparities in mental health across neighborhoods. 
    \color{black} In contrast, dynamic exposure patterns derived from mobility data, such as inter-community access to parks, show stronger correlations with mental health, especially in neighborhoods with lower Black population densities. \color{black}
    This case highlights the value of incorporating mobility-based exposure metrics into public health and urban equity research.
    
    \item \textbf{Public health and epidemiological modeling:}  The generated mobility data can be a valuable tool in public health research, particularly in understanding the spread of diseases~\cite{yuan2022activity,feng2020learning}. By capturing how individuals move within a city, public health experts can model the transmission of infectious diseases, such as COVID-19, and simulate how movement patterns influence disease spread. Additionally, by analyzing daily mobility routines, researchers can investigate how mobility affects health behaviors, such as access to healthcare facilities, physical activity levels, and dietary habits.
\end{itemize}

While WorldMove provides realistic and globally scalable synthetic mobility data, several limitations should be considered when using the dataset.
First, the urban space in each city is discretized into regular grid cells rather than administrative boundaries or semantic regions (e.g., neighborhoods or functional zones). While this allows for spatial consistency across cities, it may limit analyses that rely on fine-grained or policy-relevant geographic units.
Second, the generated trajectories cover a week, capturing typical daily and weekly mobility patterns. However, the dataset does not reflect seasonal variability, holiday-related travel behaviors, or responses to extreme events such as severe weather or lockdowns. 
Third, although the generative model is trained on a diverse set of real-world mobility sources, it inevitably inherits the sampling biases present in these datasets. These include lower-income populations, rural areas, and demographic skewness toward specific user groups (e.g., younger or tech-savvy individuals). Such biases may subtly propagate into the generated data and should be considered in downstream applications.

\color{black}

\subsection*{Practical utility of the synthetic dataset}

\begin{figure}[p]
\centering
\includegraphics[width=1.0\linewidth]{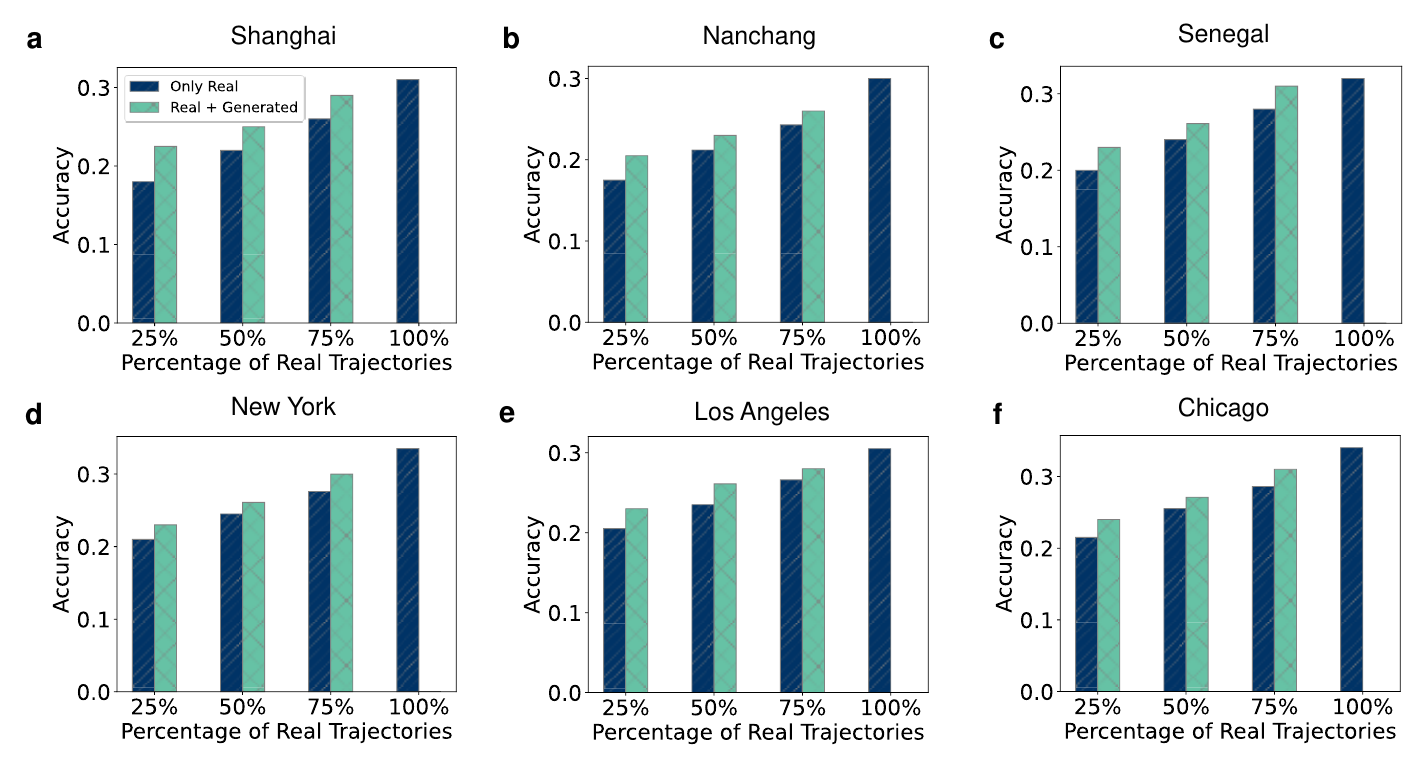}
\caption{Utility of synthetic data on mobility prediction accuracy. It compares the accuracy of mobility predictions using real data alone (blue) versus the combination of real and generated data (green) across multiple cities. As the percentage of real trajectories increases, the addition of synthetic data consistently improves prediction performance across all cities. }
\label{fig:utility}
\end{figure}

To assess the practical utility of the synthetic dataset, we tested its applicability in forecasting tasks, as shown in Figure~\ref{fig:utility}. The figure compares the accuracy of real trajectories alone versus the combination of real and synthetic data across multiple cities. The results clearly demonstrate that incorporating synthetic data alongside real data leads to an increase in accuracy. For example, in Shanghai, Nanchang, and New York, the combination of real and synthetic data consistently outperforms using only real data, especially when the percentage of real trajectories is lower (e.g., 25\%). The model shows significant improvement as more synthetic data is added, highlighting its ability to generalize and improve predictive accuracy. Similar trends are observed in other cities, such as Los Angeles, Senegal, and Chicago, where the inclusion of synthetic data boosts forecasting performance across all levels of real data availability.

These results indicate that the synthetic dataset not only captures mobility patterns effectively but also supports common data mining tasks, which are critical for traffic management and urban planning. The ability to incorporate synthetic data into predictive models provides a valuable tool for spatial-temporal analysis and ensures its practical utility in real-world applications, especially in areas with limited access to real-world data.

\color{black}



\section*{Data availability}

The synthetic mobility datasets generated in this study, covering over 1,600 cities worldwide, are openly available at the WorldMove website (\url{https://fi.ee.tsinghua.edu.cn/worldmove/}) and have also been deposited in Figshare~\cite{worldmovefigshare2024}. In addition, customized mobility data for any city can be generated using the pretrained model provided.

\section*{Code availability}
Example Python code for loading and processing the data is available at the GitHub repository: \url{https://github.com/tsinghua-fib-lab/WorldMove}, along with detailed environment requirements and installation instructions. The repository also supports generating customized mobility data for any city using the provided pretrained model.

\bibliography{reference}

\section*{Acknowledgments} 
In this work, we collect the data from multiple sources, including mobility trajectory data, WorldPop, WorldOD, and OpenStreetMap. We would like to express our gratitude to the contributors of these datasets. We also thank the advancing diffusion models, such as DDPM and DDIM,  for providing the convenient usage of their architectures as key components in our dataset construction pipeline.

\section*{Author contributions statement}

Yong Li and Yuan Yuan conceived the idea of the dataset. Yuan Yuan designed the dataset construction pipeline. Yuheng Zhang wrote the code for the generation including: downloading and processing the data from WorldPop, WorldOD, and OpenStreetMap, training the diffusion model, and generating the urban mobility. Yuan Yuan, Yuheng Zhang, Jingtao Ding, and Yong Li wrote the manuscript. All authors reviewed the manuscript.

\section*{Competing interests} (mandatory statement)

The corresponding author is responsible for providing a \href{https://www.nature.com/sdata/policies/editorial-and-publishing-policies#competing}{competing interests statement} on behalf of all authors of the paper. This statement must be included in the submitted article file.

\section*{Figures \& Tables}

\vspace{5cm}

\clearpage

\clearpage

\clearpage

\clearpage

\clearpage

\clearpage

\clearpage

\clearpage

\clearpage

\clearpage

\clearpage

\clearpage

\clearpage

\clearpage

\clearpage

\clearpage
\begin{table}[p]
\centering
\caption{Existing datasets of urban mobility.}
\label{tbl:cmopare_data}
\begin{tabular}{cccccc}
\hline
Dataset & Available &  Real-world & Type & Global & Privacy \\ \hline
GeoLife~\cite{zheng2011geolife} & \checkmark   & \checkmark & Human  & \ding{55} & \ding{55}\\ 
T-Drive~\cite{yuan2010t} & \checkmark & \checkmark & Taxi & \ding{55}  & \ding{55}\\ 
NYC Taxi~\cite{nyc_tlc_trip_data}  & \checkmark & \checkmark &  Taxi& \ding{55}  & \ding{55}\\ 
NYC-TLC~\cite{nyc_tlc_trip_data} & \checkmark  & \checkmark & Taxi& \ding{55}  & \ding{55}\\ 
Pseudo-PFLOW~\cite{kashiyama2022pseudo} & \checkmark  &  \checkmark &  Human & \ding{55}  & \ding{55}\\ 
Foursquare~\cite{yang2019revisiting}  & \checkmark &  \checkmark &   Human &  \ding{55}  & \ding{55}\\ 
SynMob~\cite{zhu2023synmob} & \ding{55} &  \ding{55} &  Taxi &  \ding{55}  & \checkmark \\
YJMob100K~\cite{yabe2024yjmob100k} & \checkmark &   \checkmark &  Human & \ding{55}  & \ding{55} \\ \hline
WorldMove (ours)  & \checkmark &   \ding{55} &  Human & \checkmark & \checkmark  \\ \hline
\end{tabular}
\end{table}

\clearpage

\begin{table}[p]
\centering
\caption{Summary of real-world mobility trajectory datasets used for training and evaluation. The "Spatial coverage" column contains the latitude (upper) and longitude (lower) ranges for each dataset.}
\label{tab:realdata}
\begin{tabular}{lcccccc}
\toprule
\textbf{dataset} & \textbf{Data source} & \textbf{Spatial coverage} & \textbf{Temporal Duration} & \textbf{\#Users} & \textbf{\#Visits} & \textbf{\#Trajectories} \\
\midrule
\multirow{2}{*}{Shanghai} & \multirow{2}{*}{Mobile operator} & [30.66, 31.88] & \multirow{2}{*}{2016.4.19-2016.4.26} & \multirow{2}{*}{29k} & \multirow{2}{*}{1749w} & \multirow{2}{*}{48w} \\
 &  & [120.86, 122.35] &  & & & \\
\multirow{2}{*}{Nanchang} & \multirow{2}{*}{Mobile operator} & [28.16, 29.18] & \multirow{2}{*}{2022.5.18-2022.5.24} & \multirow{2}{*}{6k} & \multirow{2}{*}{90w} & \multirow{2}{*}{2.6w} \\
 &  &  [115.88, 117.02] &  & & & \\
\multirow{2}{*}{Senegal} &\multirow{2}{*}{ Call Detail Record} & [12.35, 16.65] & \multirow{2}{*}{2013.1.1-2013.1.14} & \multirow{2}{*}{10w} & \multirow{2}{*}{1016w} & \multirow{2}{*}{12w} \\
 &  & [-11.37, -17.52] &  & & & \\
\multirow{2}{*}{New York} & \multirow{2}{*}{Cuebiq} & [40.47, 40.91]& \multirow{2}{*}{2016.10.1-2017.9.30} & \multirow{2}{*}{94w} & \multirow{2}{*}{30653w} & \multirow{2}{*}{1684w} \\
 &  & [-74.25, -73.70]  &  & & & \\
\multirow{2}{*}{Los Angeles} & \multirow{2}{*}{Cuebiq } & [33.70, 34.33] & \multirow{2}{*}{2016.10.1-2017.9.30} & \multirow{2}{*}{65w} & \multirow{2}{*}{35616w} & \multirow{2}{*}{1576w} \\
 &  &  [-118.66, -118.15] &  & & & \\
\multirow{2}{*}{Chicago} & \multirow{2}{*}{Cuebiq } & [41.64, 42.02] & \multirow{2}{*}{2016.10.1-2017.9.30} & \multirow{2}{*}{33w }& \multirow{2}{*}{17361w }& \multirow{2}{*}{857w} \\
 &  & [-87.94, -87.52] &  & & & \\
\bottomrule
\end{tabular}
\end{table}

\clearpage
\begin{table}[p]
\footnotesize
\centering
\caption{Evaluation of data fidelity across six cities using statistical and flow-based metrics.  
KS/JSD assess mobility pattern similarity; CPC/RMSE evaluate OD flow consistency. \color{black} This evaluation is performed under the zero-shot transfer learning setting, where the model is trained without access to target data.\color{black}
}
\vspace{-3mm}
\resizebox{1.0\textwidth}{!}{
\begin{tabularx}{\textwidth}{l|YYYY|YYYY|YY}
\toprule
\multirow{2}{*}[-0.8ex]{\textbf{Metrics}} & \multicolumn{4}{c}{\textbf{KS-Test}} & \multicolumn{4}{c}{\textbf{JSD}}
& \multicolumn{2}{c}{\textbf{Flow}} \\ 
\cmidrule(l){2-5} \cmidrule(l){6-9} \cmidrule(l){10-11}
& Radius & Distance & Duration & DailyLoc & Radius & Distance & Duration & DailyLoc & CPC & RMSE \\ 
\midrule
Shanghai     & 0.2198 & 0.2301 & 0.1347 & 0.2391
             & 0.0224 & 0.0275 & 0.0378 & 0.0245 & 0.4218 & 15.02 \\
Nanchang     & 0.1729 & 0.2273 & 0.1016 & 0.1844  
             & 0.0119 & 0.0204 & 0.0016 & 0.0213 & 0.4376 & 13.94 \\ 
Senegal      & 0.1480 & 0.2451 & 0.1795 & 0.2866
             & 0.0114 & 0.0332 & 0.0241 & 0.0431 & 0.4067 & 49.77 \\
New York     & 0.1570 & 0.1996 & 0.1315 & 0.1835
             & 0.0115 & 0.0149 & 0.0114 & 0.0224 & 0.4392 & 12.34 \\
Los Angeles  & 0.1510 & 0.2709 & 0.1871 & 0.2864 
             & 0.0015 & 0.0223 & 0.0129 & 0.0320 & 0.3884 & 17.59 \\
Chicago      & 0.1338 & 0.2654 & 0.1914 & 0.2664 
             & 0.0112 & 0.0232 & 0.0133 & 0.0427 & 0.3956 & 13.68 \\
\bottomrule
\end{tabularx}
\label{table:model_perf}
}
\end{table}

\clearpage
\begin{table}[p]
\footnotesize
\centering
\caption{Evaluation of data fidelity across six cities using statistical and flow-based metrics.  
KS/JSD assess mobility pattern similarity; CPC/RMSE evaluate OD flow consistency. \color{black}This evaluation is performed under the cross-city transfer learning setting (with target data), where the model is fine-tuned using data from the target city.\color{black}
}
\vspace{-3mm}
\resizebox{1.0\textwidth}{!}{
\begin{tabularx}{\textwidth}{l|YYYY|YYYY|YY}
\toprule
\multirow{2}{*}[-0.8ex]{\textbf{Metrics}} & \multicolumn{4}{c}{\textbf{KS-Test}} & \multicolumn{4}{c}{\textbf{JSD}}
& \multicolumn{2}{c}{\textbf{Flow}} \\ 
\cmidrule(l){2-5} \cmidrule(l){6-9} \cmidrule(l){10-11}
& Radius & Distance & Duration & DailyLoc & Radius & Distance & Duration & DailyLoc & CPC & RMSE \\ 
\midrule
Shanghai & 0.0778 & 0.1970 & 0.0967 & 0.1868 & 0.0004 & 0.0072 & 0.0045 & 0.0194 & 0.4892 & 13.98 \\
Nanchang & 0.1067 & 0.1388 & 0.0326 & 0.1200 & 0.0015 & 0.0124 & 0.0008 & 0.0187 & 0.4434 & 13.28 \\
Senegal & 0.0761 & 0.1528 & 0.0696 & 0.1842 & 0.0013 & 0.0031 & 0.0072 & 0.0126 & 0.6395 & 25.73 \\
New York & 0.0651 & 0.0949 & 0.0419 & 0.0998 & 0.0006 & 0.0072 & 0.0016 & 0.0014 & 0.6621 & 8.41 \\
Los Angeles & 0.0495 & 0.1392 & 0.1027 & 0.1739 & 0.0009 & 0.0124 & 0.0026 & 0.0247 & 0.5129 & 9.57 \\
Chicago & 0.0428 & 0.1101 & 0.0833 & 0.2184 & 0.0007 & 0.0117 & 0.0043 & 0.0119 & 0.5283 & 7.44 \\
\bottomrule
\end{tabularx}
\label{table:model_perf_insample}
}
\end{table}

\end{document}